\documentclass[fleqn,usenatbib]{mnras}

\usepackage{newtxtext,newtxmath}
\usepackage[T1]{fontenc}

\DeclareRobustCommand{\VAN}[3]{#2}
\let\VANthebibliography\thebibliography
\def\thebibliography{\DeclareRobustCommand{\VAN}[3]{##3}\VANthebibliography}

\usepackage{graphicx}	% Including figure files
\usepackage{amsmath}	% Advanced maths commands

\usepackage[normalem]{ulem}
\title[Timing the last major mergers of galaxy clusters]{Timing the last major merger of galaxy clusters with large halo sparsity}

\author[Richardson \& Corasaniti]{
{T. R. G. Richardson},$^{1}$\thanks{E-mail: thomas.richardson@obspm.fr (TR)}
{P.-S. Corasaniti},$^{1,2}$
\\
$^{1}$ Laboratoire Univers et Théories, Observatoire de Paris, Université PSL, Université de Paris, CNRS, F-92190 Meudon, France \\
$^{2}$ Sorbonne Universit\'e, CNRS, UMR 7095, Institut d'Astrophysique de Paris, 98 bis bd Arago, 75014 Paris, France\\
}

\date{Accepted XXX. Received YYY; in original form ZZZ}

\pubyear{2021}

\begin{document}
\label{firstpage}
\pagerange{\pageref{firstpage}--\pageref{lastpage}}
\maketitle

\begin{abstract}
Numerical simulations have shown that massive dark matter haloes, which today host galaxy clusters, assemble their mass over time alternating periods of quiescent accretion and phases of rapid growth associated with major merger episodes. Observations of such events in clusters can provide insights on the astrophysical processes that characterise the properties of the intra-cluster medium, as well as the gravitational processes that contribute to their assembly. It is therefore of prime interest to devise a fast and reliable way of detecting such perturbed systems. We present a novel approach to identifying and timing major mergers in clusters characterised by large values of halo sparsity. Using halo catalogues from the MultiDark-Planck2 simulation, we show that major merger events disrupt the radial mass distribution of haloes, thus leaving a distinct universal imprint on the evolution of halo sparsity over a period not exceeding two dynamical times. We exploit this feature using numerically calibrated distributions to test whether an observed galaxy cluster with given sparsity measurements has undergone a recent major merger and to eventually estimate when such an event occurred. We implement these statistical tools in a specifically developed public python library \textsc{lammas}, which we apply to the analysis of Abell 383 and Abell 2345 as test cases. Finding that, for example, Abell 2345 had a major merger about $2.1\pm 0.2$ Gyr ago. This work opens the way to detecting and timing major mergers in galaxy clusters solely through measurements of their mass at different radii.
\end{abstract}


%%%%%%%%%%%%%%%%%%%%%%%%%%%%%%%%%%%%%%%%%%%%%%%%%%

%%%%%%%%%%%%%%%%% BODY OF PAPER %%%%%%%%%%%%%%%%%%

\section{Introduction}\label{sec:introduction}
Galaxy clusters are the ultimate result of the hierarchical bottom-up process of cosmic structure formation. Hosted in massive dark matter haloes that formed through subsequent phases of mass accretion and mergers, galaxy clusters carry information on the underlying cosmological scenario as well as the astrophysical processes that shape the properties of the intra-cluster medium (ICM) \citep[for a review, see e.g.][]{2005RvMP...77..207V,2011ARA&A..49..409A,2012ARA&A..50..353K}. 

Being at the top of the pyramid of cosmic structures, galaxy clusters are mostly found in the late-time universe. These can be observed using a variety a techniques that probe either the distribution of the hot intra-cluster gas through its X-ray emission \citep[see e.g.][]{2005ApJ...628..655V,2010MNRAS.407...83E,2016A&A...592A...1P,2021A&A...650A.104C}, the scattering of the Cosmic Microwave Background radiation (CMB) due to the Sunyaev-Zeldovich effect \citep[see e.g.][]{2009ApJ...701...32S,2013ApJ...765...67M,2013ApJ...763..127R,2014A&A...571A..29P,2015ApJS..216...27B}, through measurement of galaxy overdensities or the gravitational lensing effect caused by the cluster's gravitational mass on background sources \citep{2016ApJS..224....1R,2019MNRAS.485..498M,2011ApJ...738...41U,2012ApJS..199...25P}.

The mass distribution of galaxy clusters primarily depends on the dynamical state of the system. Observations of relaxed clusters have shown that the matter density profile at large radii is consistent with the universal Navarro-Frenk-White profile \citep[NFW,][]{NFW1997}, while deviations have been found in the inner regions \citep[][]{2013ApJ...765...24N,2017ApJ...851...81A,2017ApJ...843..148C,2020A&A...637A..34S}. In relaxed systems, the gas falls in the dark matter dominated gravitational potential and thermalises through the propagation of shock waves. This sets the gas in a hydrostatic equilibrium (HE) that is entirely controlled by gravity. Henceforth, aside astrophysical processes affecting the baryon distribution in the cluster core, the thermodynamic properties of the outer ICM are expected to be self-similar \citep[see e.g.][]{2019A&A...621A..39E,2019A&A...621A..41G,2021ApJ...910...14G}. This is not the case of clusters undergoing major mergers for which the virial equilibrium is strongly altered \citep[see e.g.][]{2016ApJ...827..112B}. Such systems exhibit deviations from self-similarity such that scaling relations between the ICM temperature, the cluster mass and X-ray luminosity differ from that of relaxed clusters \citep[see e.g.][]{2009MNRAS.399..410P,2011ApJ...729...45R,2019MNRAS.490.2380C}. 

A direct consequence of merger events is that the mass estimates inferred assuming the HE hypothesis or through scaling relations may be biased. This may induce systematic errors on cosmological analyses that rely upon accurate cluster mass measurements. On the other hand, merging clusters can provide a unique opportunity to investigate the physics of the ICM \citep{2007PhR...443....1M,2016JPlPh..82c5301Z} and test the dark matter paradigm \citep[as in the case of the Bullet Cluster][]{2004ApJ...604..596C,2004ApJ...606..819M}. This underlies the importance of identifying merging events in large cluster survey catalogues.

The identification of unrelaxed clusters relies upon a variety of proxies specifically defined for each type of observations \citep[for a review see e.g.][]{2016FrASS...2....7M}. As an example, the detection of radio haloes and relics in clusters is usually associated with the presence of mergers. Similarly, the offset between the position of the brightest central galaxy and the peak of the X-ray surface brightness, or the centroid of the SZ signal are used as proxy of merger events. This is because the merging process alters differently the distribution of the various matter constituents of the cluster.

The growth of dark matter haloes through cosmic time has been investigated extensively in a vast literature using results from N-body simulations. \citet{2003MNRAS.339...12Z} found that haloes build up their mass through an initial phase of fast accretion followed by a slow one. \citet{2007MNRAS.379..689L} have shown that the during the fast-accretion phase, the mass assembly occurs primarily through major mergers, that is mergers in which the mass of the less massive progenitor is at least one third of the more massive one. Moreover, they found that the greater the mass of the halo the later the time when the major merger occurred. In contrast, slow accretion is a quiescent phase dominated by minor mergers. Subsequent studies have mostly focused on the relation between the halo mass accretion history and the concentration parameter of the NFW profile \citep[see e.g.][]{2007MNRAS.381.1450N,2009ApJ...707..354Z,2012MNRAS.427.1322L,2016MNRAS.460.1214L,2017MNRAS.466.3834L,2019MNRAS.485.1906R}. Recently, \citet{Wang2020} have shown that major mergers have a universal impact on the evolution of the median concentration. In particular, after a large initial response, in which the concentration undergoes a large excursion, the halo recovers a more quiescent dynamical state within a few dynamical times. Surprisingly, the authors have also found that even minor mergers can have a non-negligible impact on the mass distribution of haloes, contributing to the scatter of the concentration parameter. 

The use of concentration as a proxy of galaxy mergers is nevertheless challenging for multiple reasons. Firstly, the concentration exhibits a large scatter across the merger phase and the value inferred from the analysis of galaxy cluster observations may be sensitive to the quality of the NFW-fit. Secondly, astrophysical processes may alter the mass distribution in the inner region of the halo, thus resulting in values of the concentration that differ from those estimated from N-body simulations \citep[see e.g.][]{2010MNRAS.406..434M,2011MNRAS.416.2539K}, which could be especially the case for merging clusters. 

Alternatively, a non-parametric approach to characterise the mass distribution in haloes has been proposed by \citet{Balmes2014} in terms of simple mass ratios, dubbed halo {\it sparsity}:
\begin{equation}\label{sparsdef}
s_{\Delta_1,\Delta_2} = \frac{M_{\Delta_1}}{M_{\Delta_2}},
\end{equation}  
where $M_{\Delta_1}$ and $M_{\Delta_2}$ are the masses within spheres enclosing respectively the overdensity $\Delta_1$ and $\Delta_2$ (with $\Delta_1<\Delta_2$) in units of the critical density (or equivalently the background density). This statistics presents a number of interesting properties that overcome many of the limitations that concern the concentration parameter. First of all, the sparsity can be estimated directly from cluster mass estimates without having to rely on the assumption of a specific parametric profile, such as the NFW profile. Secondly, for any given choice of $\Delta_1$ and $\Delta_2$, the sparsity is  found to be weakly dependent on the overall halo mass with a much reduced scatter than the concentration \citep{Balmes2014,Corasaniti2018,Corasaniti2019}. Thirdly, these mass ratios retain cosmological information encoded in the mass profile, thus providing an independent cosmological proxy. Finally, the halo ensemble average sparsity can be predicted from prior knowledge of the halo mass functions at the overdensities of interests, which allows to infer cosmological parameter constraints from cluster sparsity measurements \citep[see e.g.][]{Corasaniti2018,Corasaniti2021}.  

As haloes grow from inside out such that newly accreted mass is redistributed in concentric shells within a few dynamical times \citep[see e.g.][for a review]{2011MNRAS.413.1373W,2011AdAst2011E...6T}, it is natural to expect that major mergers can significantly disrupt the onion structure of haloes and result in values of the sparsity that significantly differ from those of the population of haloes that have had sufficient time to rearrange their mass distribution and reach the virial equilibrium. 

Here, we perform a thorough analysis of the relation between halo sparsity and the halo mass accretion history using numerical halo catalogues from large volume high-resolution N-body simulations. We show that haloes which undergo a major merger in their recent history form a distinct population of haloes characterised by large sparsity values. Quite importantly, we are able to fully characterise the statistical distributions of such populations in terms of the halo sparsity and the time of their last major merger. Thus, building upon these results, we have developed a statistical tool which uses cluster sparsity measurements to test whether a galaxy cluster has undergone a recent major merger and if so when such event took place.

The paper is organised as follows. In Section~\ref{halocat} we describe the numerical halo catalogues used in the analysis, while in Section~\ref{sparsmah} we present the results of the study of the relation between halo sparsity and major mergers. In Section~\ref{calistat} we present the statistical tests devised to identify the imprint of mergers in galaxy clusters and in  discuss the statistical estimation of the major merger epoch from sparsity measurements. In Section~\ref{cosmo_imp} we discuss the implications of these results regarding cosmological parameter estimation studies using halo sparsity. In Section~\ref{testcase} we validate our approach using similar data, assess its robustness to observational biasses and describe the application of our methodology to the analysis of known galaxy clusters. Finally, in Section~\ref{conclu} we discuss the conclusions.

\section{Numerical Simulation Dataset}\label{halocat}
\subsection{N-body Halo catalogues}
We use N-body halo catalogues from the MultiDark-Planck2 (MDPL2) simulation \citep{Klypin2016} which consists of $3840^3$ particles in $(1 \,h^{-1}\,\textrm{Gpc})^3$ comoving volume (corresponding to a particle mass resolution of $m_p=1.51\cdot 10^{9}\,h^{-1} \text{M}_{\odot}$) of a flat $\Lambda$CDM cosmology run with the \textsc{Gadget-2}\footnote{\href{https://wwwmpa.mpa-garching.mpg.de/gadget/}{https://wwwmpa.mpa-garching.mpg.de/gadget/}} code \citep{2005MNRAS.364.1105S}. The cosmological parameters have been set to the values of the \textit{Planck} cosmological analysis of the Cosmic Microwave Background (CMB) anisotropy power spectra \citep{2014A&A...571A..16P}: $\Omega_m=0.3071$, $\Omega_b=0.0482$, $h=0.6776$, $n_s=0.96$ and $\sigma_8=0.8228$. Halo catalogues and merger trees at each redshift snapshot were generated using the friend-of-friend (FoF) halo finder code \textsc{rockstar}\footnote{\href{https://code.google.com/archive/p/rockstar/}{https://code.google.com/archive/p/rockstar/}} \citep{Behroozi2013a,Behroozi2013b}. We consider the default set up with the detected haloes consisting of gravitationally bound particles only. We specifically focus on haloes in the mass range of galaxy groups and clusters corresponding to $M_{200\text{c}}>10^{13}\,h^{-1} \text{M}_{\odot}$. 

For each halo in the MDPL2 catalogues we build a dataset containing the following set of variables: the halo masses $M_{200\text{c}}$, $M_{500\text{c}}$ and $M_{2500\text{c}}$ estimated from the number of N-body particles within spheres enclosing overdensities $\Delta=200,500$ and $2500$ (in units of the critical density) respectively; the scale radius, $r_s$, of the best-fitting NFW profile; the virial radius, $r_{\rm vir}$; the ratio of the kinetic to the potential energy, $K/U$; the offset of the density peak from the average particle position, $x_{\rm off}$; and the scale factor (redshift) of the last major merger, $a_{\rm LMM}$ ($z_{\rm LMM}$). From these variables we additionally compute the following set of quantities: the halo sparsities $s_{200,500}$, $s_{200,2500}$ and $s_{500,2500}$; the offset in units of the virial radius, $\Delta_r=x_{\rm off}/r_{\rm vir}$, and the concentration parameter of the best-fit NFW profile, $c_{200\text{c}}=r_{200\text{c}}/r_s$, with $r_{200\text{c}}$ being the radius enclosing an overdensity $\Delta=200$ (in units of the critical density). In our analysis we also use the mass accretion history of MDPL2 haloes.

In addition to the MDPL2 catalogues, we also use data from the Uchuu simulations \citep{Ishiyama2021}, which cover a larger cosmic volume with higher mass resolution. We use these catalogues to calibrate the sparsity statistics that provides the base for practical applications of halo sparsity measurements as cosmic chronometers of galaxy cluster mergers. The Uchuu simulation suite consists of N-body simulations of a flat $\Lambda$CDM model realised with \textsc{GreeM} code \citep{2009PASJ...61.1319I,2012arXiv1211.4406I} with cosmological parameters set to the values of a later \textit{Planck}-CMB cosmological analysis  \citep{2016A&A...594A..13P}: $\Omega_m=0.3089$, $\Omega_b=0.0486$, $h=0.6774$, $n_s=0.9667$ and $\sigma_8=0.8159$. In particular, we use the halo catalogues from the $(2\,\textrm{Gpc}\,h^{-1})^3$ comoving volume simulation with $12800^3$ particles (corresponding to a particle mass resolution of $m_p=3.27\cdot 10^{8}\,h^{-1}\text{M}_{\odot}$) that, as for MDPL2, were also generated using the \textsc{rockstar} halo finder.

It is important to stress that the major merger epoch to which we refer in this work is that defined by the \textsc{rockstar} halo finder, that is the time when the particles of the merging halo and those of the parent one are within the same iso-density contour in phase-space. Hence, this should not be confused with the first core-passage time usually estimated in Bullet-like clusters.

\begin{table}
\centering
\caption{Characteristics of the selected halo samples at $z=0,0.2,0.4$ and $0.6$ (columns from left to right). Quoted in the rows are the number of haloes in the samples and the redshift of the last major merger $z_{\rm LMM}$ used to select the haloes for each sample.}
\begin{tabular}{ccccc}
\hline
\hline
& \multicolumn{4}{c}{Merging Halo Sample ($T>-1/2$)} \\ 
\hline
\hline
 & $z=0.0$ & $z=0.2$ & $z=0.4$ & $z=0.6$ \\
 \hline
$\#$-haloes & $23164$ & $28506$ & $31903$ & $32769$ \\
$z_{\rm LMM}$ & $<0.113$ & $<0.326$ & $<0.540$ & $<0.754$ \\
\hline
\hline
& \multicolumn{4}{c}{Quiescent Halo Sample ($T<-4$)} \\ 
\hline
\hline
 & $z=0.0$ & $z=0.2$ & $z=0.4$ & $z=0.6$ \\
 \hline
$\#$-haloes & $199853$ & $169490$ & $140464$ & $113829$ \\
$z_{\rm LMM}$ & $>1.15$ & $>1.50$ & $>1.86$ & $>2.22$ \\
\hline
\end{tabular}
\label{tab:samples}
\end{table}

\subsection{Halo Sample Selection}\label{haloeselection}
We aim to study the impact of merger events on the halo mass profile. To this purpose we focus on haloes which undergo their last major merger at different epochs. In such a case, it is convenient to introduce a time variable that characterises the backward time interval between the redshift $z$ (scale factor $a$) at which a halo is investigated and that of its last major merger $z_{\rm LMM}$ ($a_{\rm LMM}$) in units of the dynamical time \citep{Jiang2016, Wang2020},
\begin{equation}\label{backwardtime}
T(z|z_\text{LMM})= \frac{\sqrt{2}}{\pi}\int_{z_{\text{LMM}}}^{z}\frac{\sqrt{\Delta_\text{vir}(z)}}{z+1}dz,
\end{equation}
where $\Delta_{\rm vir}(z)$ is the virial overdensity, which we estimate using the spherical collapse model approximated formula $\Delta_{\rm vir}(z)=18\pi^2+82[\Omega_m(z)-1]-39[\Omega_m(z)-1]^2$ \citep{Bryan1998}. Hence, one has $T=0$ for haloes which undergo a major merger at the time they are investigated (i.e. $z_{\rm LMM}=z$), and $T<0$ for haloes that had their last major merger at earlier times (i.e. $z_{\rm LMM}>z$). Notice that the definition used here differs by a minus sign from that of \citet{Wang2020}, where the authors have found that merging haloes recover a quiescent state within $|T| \sim 2$ dynamical times. 

In Section~\ref{sparsprof}, we investigate the differences in halo mass profile between merging haloes and quiescent ones, to maximise the differences we select haloes samples as following:
\begin{itemize}
\item {\it Merging haloes}: a sample of haloes that are at less than one half the dynamical time since their last major merger ($T> -1/2$), and therefore still in the process of rearranging their mass distribution;
\item {\it Quiescent haloes}: a sample of haloes for which their last major merger occurred far in the past ($T\le -4)$, thus they had sufficient time to rearrange their mass distribution to an equilibrium state; 
\end{itemize}

In the case of the $z=0$ catalogue, the sample of merging haloes with $T>-1/2$ consists of all haloes for which their last major merger as tagged by the \textsc{rockstar} algorithm occurred at $a_{\rm LMM}>0.897$ ($z_{\rm LMM}<0.115$), while the samples of quiescent
%and relaxed quiescent
haloes with $T\le -4$ in the same catalogue are characterised by a last major merger at $a_{\rm LMM}<0.464$ ($z_{\rm LMM}>1.155$). In order to study the redshift dependence, we perform a similar selection for the catalogues at $z=0.2,0.4$ and $0.6$ respectively. In Table~\ref{tab:samples} we quote the characteristics of the different samples selected in the various catalogues.

\begin{figure*}
    \centering
    \includegraphics[width=.8\linewidth]{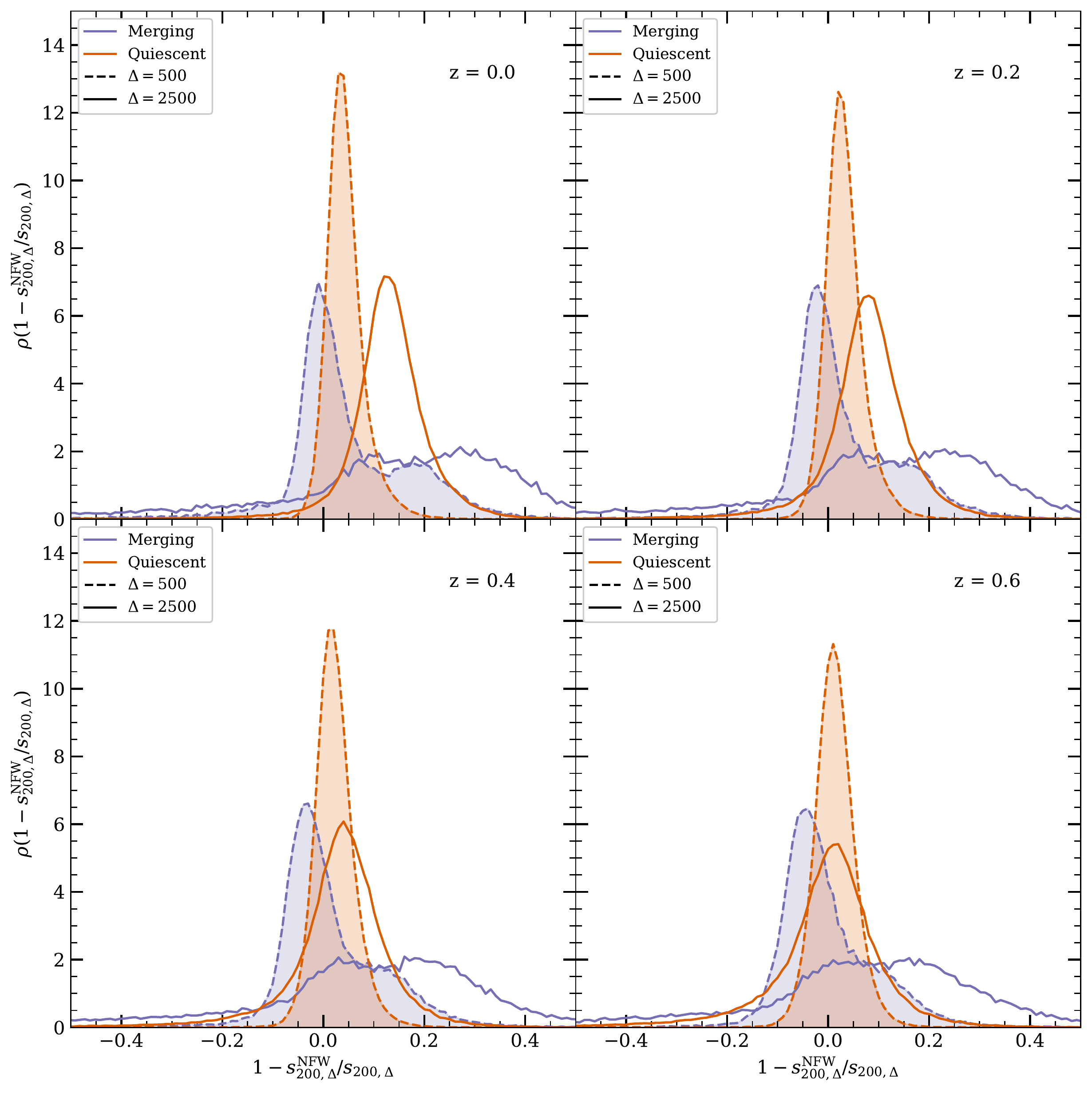}
    \caption{Distribution of the relative deviations of individual halo sparsities with respect to the expected NFW value for $\delta_{200,500}=1-s^{\rm NFW}_{200,500}/s_{200,500}$ (dashed lines) and $\delta_{200,2500}=1-s^{\rm NFW}_{200,2500}/s_{200,2500}$ (solid lines) in the case of the merging (blue lines) and quiescent (orange lines) haloes at $z=0.0$ (top left panel), $0.2$ (top right panel), $0.4$ (bottom left panel) and $0.6$ (bottom right panel) respectively}
    \label{fig:relative_spars_conc}
\end{figure*}

\section{Halo Sparsity \& Major Mergers}\label{sparsmah}
\subsection{Halo Sparsity Profile}\label{sparsprof}
Here, we seek to investigate the halo mass profile of haloes undergoing a major merger as traced by halo sparsity and evaluate to which extent the NFW profile can account for the estimated sparsities at different overdensities. To this purpose, we compute for each halo in selected samples the halo sparsities $s_{200,500}$ and $s_{200,2500}$ from their SOD estimated masses, as well as the values obtained assuming the NFW profile using the best-fit concentration parameter $c_{200\text{c}}$, which we denote as $s^{\rm NFW}_{200,500}$ and $s^{\rm NFW}_{200,2500}$ respectively. These can be inferred from the sparsity-concentration relation \citep{Balmes2014}:
\begin{equation}
x^3_{\Delta}\frac{\Delta}{200}=\frac{\ln{(1+c_{200\text{c}}x_{\Delta})}-\frac{c_{200\text{c}}x_{\Delta}}{1+c_{200\text{c}}x_{\Delta}}}{\ln{(1+c_{200\text{c}})}-\frac{c_{200\text{c}}}{1+c_{200\text{c}}}},\label{sparconc}
\end{equation}
where $x_{\Delta}=r_{\Delta}/r_{200\text{c}}$ with $r_{\Delta}$ being the radius enclosing $\Delta$ times the critical density. Hence, for any value of $\Delta$ and given the value of $c_{200\text{c}}$ for which the NFW-profile best-fit that of the halo of interest, we can solve Eq.~(\ref{sparconc}) numerically to obtain $x_{\Delta}$ and then derive the value of the NFW halo sparsity given by:
\begin{equation}
s^{\rm NFW}_{200,\Delta}=\frac{200}{\Delta}x_{\Delta}^{-3}.
\end{equation}
It is worth emphasising that such relation holds true only for haloes whose density profile is well described by the NFW formula. In such a case the higher the concentration the smaller the value of sparsity, and inversely the lower the concentration the higher the sparsity. Because of this, the mass ratio defined by Eq.~(\ref{sparsdef}) provides information on the level of sparseness of the mass distribution within haloes, that justifies being dubbed as halo sparsity. Notice that from Eq.~(\ref{sparconc}) we can compute $s_{200,\Delta}$ for any $\Delta>200$, and this is sufficient to estimate the sparsity at any other pair of overdensities $\Delta_1\ne\Delta_2>200$ as given by $s_{\Delta_1,\Delta_2}=s_{200,\Delta_1}/s_{200,\Delta_2}$. Haloes whose mass profile deviates from the NFW prediction will have sparsity values that differ from that given by Eq.~(\ref{sparconc}).

This is emphasised in Fig.~\ref{fig:relative_spars_conc}, where we plot the distribution of the relative deviations of individual halo sparsities with respect to the expected NFW value for $\delta_{200,500}=1-s^{\rm NFW}_{200,500}/s_{200,500}$ (dashed lines) and $\delta_{200,2500}=1-s^{\rm NFW}_{200,2500}/s_{200,2500}$ (solid lines) in the case of the merging (blue lines) and quiescent (orange lines) haloes at $z=0.0,0.2,0.4$ and $0.6$ respectively. We can see that for quiescent haloes the distributions are nearly Gaussian. More specifically, in the case $\delta_{200,500}$ we can see that the distribution has a narrow scatter with a peak that is centred at the origin at $z=0.6$, and slightly shifts toward positive values at smaller redshifts with a maximal displacement at $z=0$. This corresponds to an average bias of the NFW-estimated sparsity $s^{\rm NFW}_{200,500}$ of order $\sim 4\%$ at $z=0$. A similar trend occurs for the distribution of $\delta_{200,2500}$, though with a larger scatter and a larger shift in the location of the peak of the distribution at $z=0$, which corresponds to an average bias of $s^{\rm NFW}_{200,2500}$ of order $\sim 14\%$ at $z=0$. Such systematic differences are indicative of the limits of the NFW-profile in reproducing the halo mass profile of haloes both in the outskirt regions and the inner ones. Moreover, the redshift trend is consistent with the results of the analysis of the mass profile of stacked haloes presented in \citep[][]{2018ApJ...859...55C}, which shows that the NFW-profile better reproduce the halo mass distribution at $z=3$ than at $z=0$ (see top panels of their Fig.~8). Very different is the case of the merging halo sample, for which we find the distribution of $\delta_{200,500}$ and $\delta_{200,2500}$ to be highly non-Gaussian and irregular. In particular, in the case of $\delta_{200,500}$ we find the distribution to be characterised by a main peak located near the origin with a very heavy tail up to relative differences of order $20\%$. The effect is even more dramatic for $\delta_{200,2500}$, in which case the distribution looses the main peak and become nearly bimodal, while being shifted over a positive range of values that extend up to relative variations up $\sim 40\%$. Overall this suggests that  sparsity provides a more reliable proxy of the halo mass profile than that inferred from the NFW concentration.

\begin{figure}
    \centering
    \includegraphics[width = \linewidth]{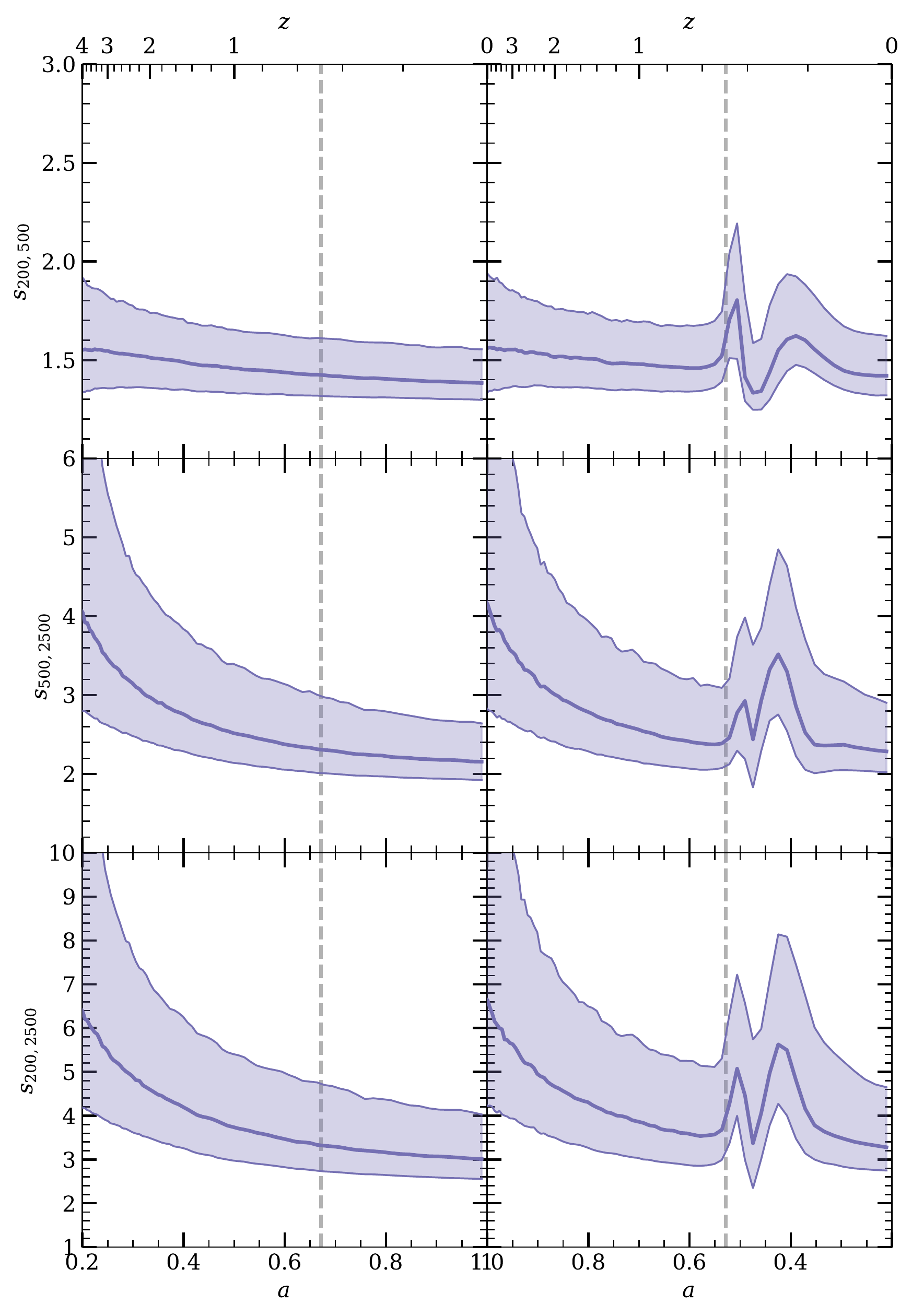}
    \caption{Evolution with scale factor $a$ (redshift $z$) of the median sparsity $s_{200,500}$ (top panels), $s_{500,2500}$ (middle panels) and $s_{200,2500}$ (bottom panels) for a sample of $10^4$ randomly selected haloes from the MDPL2 halo catalogue at $z=0$ and the sample of all haloes with a last major merger event at $a_{\rm LMM} = 0.67$ (right panels). The solid lines corresponds to the median sparsity computed from the mass accretion histories of the individual haloes, while the shaded area corresponds to the $68\%$ region around the median.}
    \label{fig:sparsity_histories_1}
\end{figure}
\begin{figure}
    \centering
    \includegraphics[width = 0.8\linewidth]{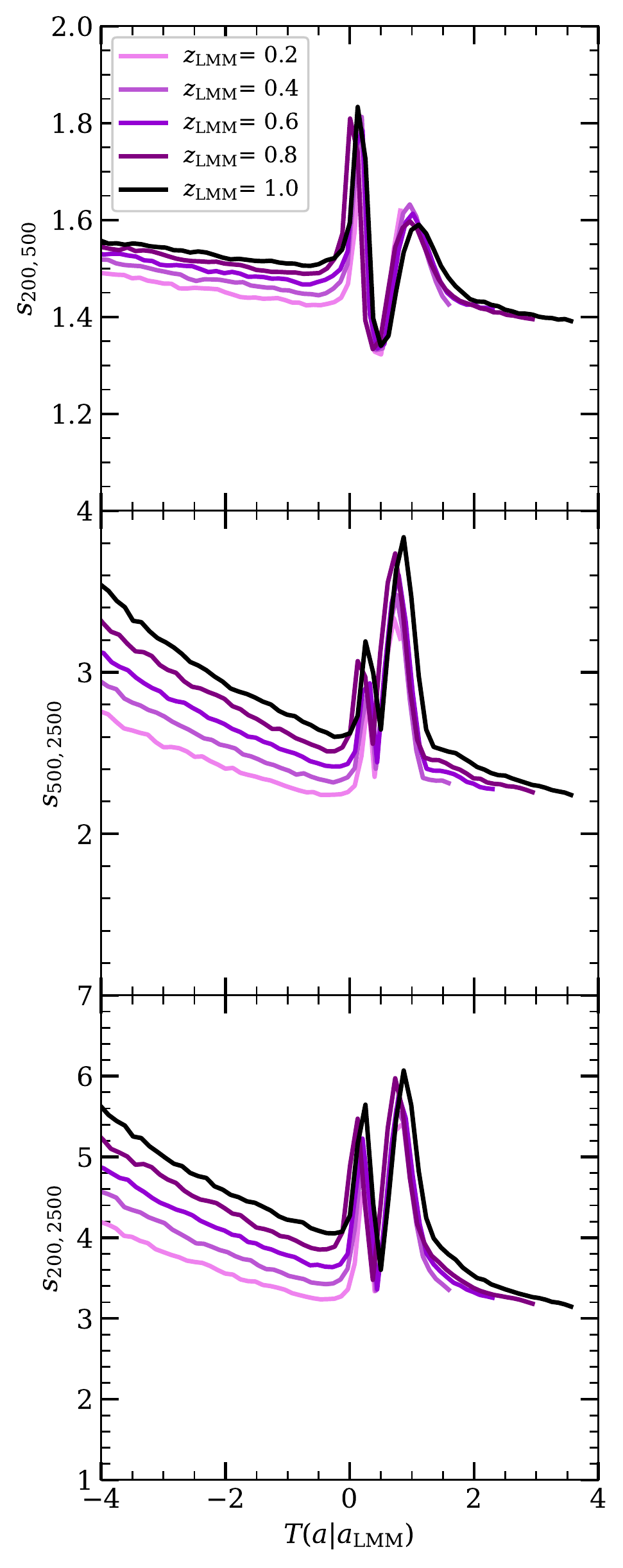}
    \caption{Median sparsity histories as function of the backward time interval since the major merger events $T$ (in units of dynamical time) for halo samples from the MDPL2 catalogue at $z=0$ with different last major merger redshifts $z_{\rm LMM}=0.2,0.4,0.6,0.8,0.8$ and $1$ (curves from bottom to top). Notice that the backward time interval used here differ by a minus sign from that given by Eq.~(\ref{backwardtime}) to be consistent with the definition by \citet{Wang2020}.}
    \label{fig:sparsity_histories_2}
\end{figure}

\subsection{Halo Sparsity Evolution}
Differently from the previous analysis, we now investigate the evolution of the halo mass profile as traced by halo sparsity, which we reconstruct from the mass accretion histories of the haloes in the MDPL2 catalogue at $z=0$. In Fig.~\ref{fig:sparsity_histories_1}, we plot the median sparsity evolution of $s_{200,500}$ (top panel), $s_{500,2500}$ (middle panel) and $s_{200,2500}$ (bottom panel) as function of the scale factor. In the left panels we shown the case of a sample of $10^4$ randomly selected haloes, thus behaving as quiescent haloes in the redshift range considered, while in the right panels we plot the evolution of the sparsity of all haloes in the $z=0$ catalogue undergoing a major merger at $a_{\rm LMM}=0.67$. The shaded area corresponds to the $68\%$ sparsity excursion around the median, while the vertical dashed line marks the value of the scale factor of the last major merger.

It is worth remarking taht the sparsity provides us with an estimate of the fraction of mass in a shell of radii $R_{\Delta_1}$ and $R_{\Delta_2}$ relative to the mass enclosed in the inner radius $R_{\Delta_2}$, i.e. Eq.~(\ref{sparsdef}) can be rewritten $s_{\Delta_1,\Delta_2}=\Delta{M}/M_{\Delta_2}+1$. As such $s_{200,500}$ is a more sensitive probe of the mass distribution in the external region of the halo, while $s_{500,2500}$ and $s_{200,2500}$ are more sensitive to the inner part of the halo. 

As we can see from Fig.~\ref{fig:sparsity_histories_1}, the evolution of the sparsity of merging haloes matches that of the quiescent sample before the major merger event. In particular, during the quiescent phase of evolution, we notice that $s_{200,500}$ remains nearly constant, while $s_{500,2500}$ and $s_{200,2500}$ are decreasing functions of the scale factor. This is consistent with the picture that haloes grow from inside out, with the mass in the inner region (in our case $M_{2500\text{c}}$) increasing relative to that in the external shell ($\Delta{M}=M_{\Delta_1}-M_{2500\text{c}}$, with $\Delta_1=200$ and $500$ in units of critical density), thus effectively reducing the value of the sparsity. This effect is compensated on $s_{200,500}$, thus resulting in a constant evolution. 
We can see that the onset of the major merger event induce a pulse-like response in the evolution of halo sparsities at the different overdensities with respect to the quiescent evolution. These trends are consistent with the evolution of the median concentration during major mergers found in \citet{Wang2020}, in which the concentration rapidly drops to a minimum before bouncing again. Here, the evolution of the sparsity allows to follow how the merger alters the mass profile of the halo throughout the merging process. In fact, we may notice that the sparsities rapidly increase to a maximum, suggesting the arrival of the merger in the external region of the parent halo, which increases the mass $\Delta{M}$ in the outer shell relative to the inner mass. Then, the sparsities decrease to a minimum, indicating that the merged mass has reached the inner region, after which the sparsities increases to a second maximum that indicates that the merged mass has been redistributed outside the $R_{2500\text{c}}$ radius. However, notice that in the case of $s_{200,2500}$ and $s_{500,2500}$ the second peak is more pronounced than the first one, while the opposite occurs for $s_{200,500}$, which suggests that the accreted mass remains confined within $R_{500\text{c}}$. Afterwards, a quiescent state of evolution is recovered.

In Fig.~\ref{fig:sparsity_histories_2} we plot the median sparsities of haloes in the MDPL2 catalogue at $z=0$ that are characterised by different major merger redshifts $z_{\rm LMM}$ as function of the backward interval of time $T$ (in units of the dyanmical time) since the last major merger. Notice that $T$ used in this plot differs by a minus sign from that given by Eq.~(\ref{backwardtime}) to conform to the definition by \citet{Wang2020}. We can see that after the onset of the major merger (at $T\ge 0$), the different curves superimpose on one another, indicating that the imprint of the major merger on the profile of haloes is universal, producing the same pulse-like feature on the evolution of the halo sparsity. Furthermore, all haloes recover a quiescent evolution within two dynamical times, i.e. for $T\ge 2$. Conversely, on smaller time scale $T<2$, haloes are still perturbed by the major merger event. These result are consistent with the findings of \citet{Wang2020}, who have shown that the impact of mergers on the median concentration of haloes leads to a time pattern that is universal and also dissipates within two dynamical times. Notice, that this distinct pattern due to the major merger is the result of gravitational interactions only. Hence, it is possible that such a feature may be sensitive to the underlying theory of gravity or the physics of dark matter particles.

As we will see next, the universality of the pulse-like imprint of the merger event on the evolution of the halo sparsity, as well as its limited duration in time, have quite important consequences, since these leave a distinct feature on the statistical distribution of sparsity values, which can be exploited to use sparsity measurements as a time proxy of major mergers in clusters.

\begin{figure*}
    \centering
    \includegraphics[width = 0.8\linewidth]{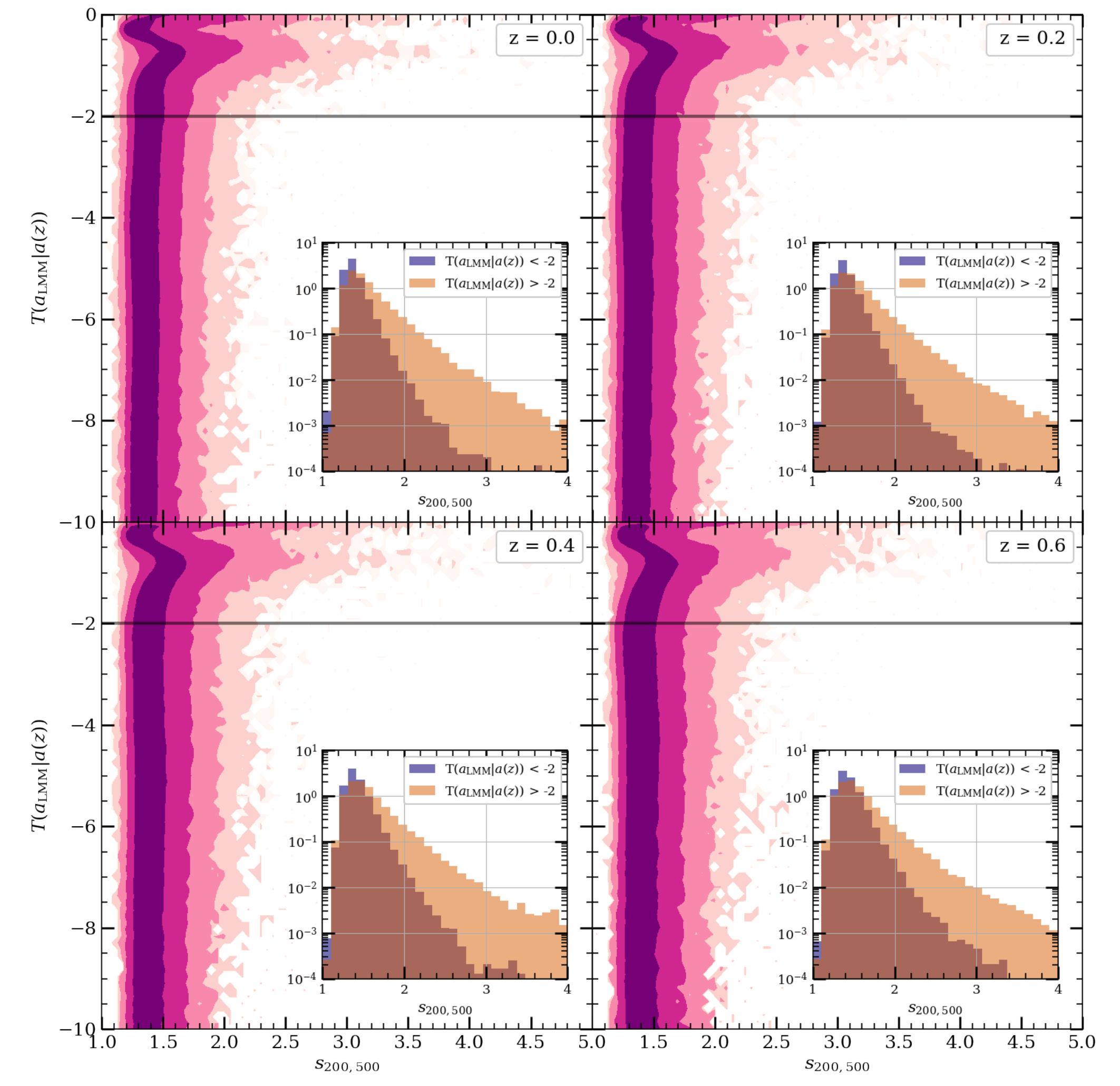}
    \caption{\label{fig:s_almm} Iso-probability contours of the joint probability distribution in the $s_{200,500}-T$ plane for the haloes from the MDPL2 catalogues at $z=0.0,0.2,0.4$ and $0.6$ respectively. The solid horizontal line marks the value $T=-2$. The inset plots show that marginal probability distribution for haloes with $T>-2$ (blue histograms) and $T<-2$ (beige histograms) respectively.}
    \label{fig:sva}
\end{figure*}

\subsection{Halo Sparsity Distribution}
We have seen that the sparsity of different haloes evolves following the same pattern after the onset of the major merger, such that the universal imprint of the merger event is best highlighted in terms of the backward interval time $T$. Hence, we aim to investigate the joint statistical distribution of halo sparsity values for haloes characterised by different time $T$ since their last major merger in the MDPL2 catalogues at different redshift. Here, we revert to the definition of $T$ given by Eq.~(\ref{backwardtime}), where the time interval is measured relative to the time the haloes are investigated, that is the redshift $z$ of the halo catalog. Hence, $T=0$ for haloes undergoing a major merger at $z_{\rm LMM}=z$ and $T<0$ for those with $z_{\rm LMM}>z$.

For conciseness, here we only describe the features of the joint distribution $p(s_{200,500},T)$ shown in Fig.~\ref{fig:s_almm} in the form of iso-probability contours in the $s_{200,500}-T$ plane at $z=0$ (top left panel), $0.2$ (top right panel), $0.4$ (bottom left panel) and $0.6$ (bottom right panel). We find a similar structure of the distributions at other redshift snapshots and for the halo sparsities $s_{200,2500}$ and $s_{500,2500}$. In each panel the horizontal solid line marks the characteristic time interval $|T|=2$. As shown by the analysis of the evolution of the halo sparsity, haloes with $|T|>2$ have recovered a quiescent state, while those with $|T|<2$ are still undergoing the merging process. The marginal conditional probability distributions $p(s_{200,500}|T<-2)$ and $p(s_{200,500}|T>-2)$ are shown in the inset plot. 

Firstly, we may notice that the joint probability distribution has a universal structure, that is the same at different redshift snapshots. Moreover, it is characterised by two distinct regions. The region with $T\le -2$, that corresponds to haloes which are several dynamical times away since their last major merger event ($|T|\ge 2$), as such they are in a quiescent state of evolution of the sparsity; and a region with $-2<T<0$, corresponding to haloes that are still in the merging processes ($|T|<2$). In the former case, the pdf has a rather regular structure that is independent of $T$, while in the latter case the pdf has an altered structure with a pulse-like feature shifted toward higher sparsity values. The presence of such a feature is consistent with the evolution of the median sparsity inferred from the halo mass accretion histories previously discussed. This is because among the haloes observed at a given redshift snapshot, those which are within two dynamical times from the major merger event are perturbed, thus exhibiting sparsity values that are distributed around the median shown in Fig.~\ref{fig:sparsity_histories_2}. In contrast, those which are more than two dynamical times since their last major merger had time to redistributed the accreted mass and are in a quiescent state, causing a regular structure of the pdf. From the inset plots, we can see that these two regions identify two distinct population of haloes, quiescent haloes with $T\le -2$ and merging (or perturbed) ones with $-2<T<0$ characterised by a stiff tail toward large sparsity values and that largely contributes to the overall scatter of the halo sparsity of the entire halo ensemble. It is worth stressing that the choice of $T=2$ as threshold to differentiate between the quiescent haloes and the perturbed ones at a given redshift snapshot is not arbitrary, since it is the most conservative value of the dynamical time above which haloes that have undergone a major merger recover a quiescent evolution of their halo mass profile as shown in Fig.~\ref{fig:sparsity_histories_2}.

Now, the fact that two populations of haloes have different probability distribution functions suggests that measurements of cluster sparsity can be used to identify perturbed systems that have undergone a major merger.

\section{Identifying Galaxy Cluster Major Mergers}\label{calistat}
Given the universal structure of the probability distributions characterising merging haloes and quiescent ones, we can use the numerical halo catalogues to calibrate their statistics at different redshifts and test whether a cluster with a single or multiple sparsity measurements has had a major merger in its recent mass assembly history. 

In the light of these observations, we first design a method to assess whether a cluster has or hasn't been recently perturbed by a major merger. To do this we design a binary test, as defined in detection theory \citep[see e.g.][]{kay1998fundamentals}, to differentiate between the different cases. Formally, this translates into defining two hypotheses denoted as $\mathcal{H}_0$, the null hypothesis and, $\mathcal{H}_1$, the alternate hypothesis. In our case these are, $\mathcal{H}_0$: \textit{The halo has not been recently perturbed} and $\mathcal{H}_1$: \textit{The halo has undergone a recent major merger}. Formally the distinction between the two is given in terms of the backward time interval $T$,
\begin{equation}
    \begin{cases}
    \mathcal{H}_0:\; T(a_\text{LMM}|a(z)) < -2\\
    \mathcal{H}_1:\; T(a_\text{LMM}|a(z)) \geq -2
    \end{cases}
    \label{eq:hypothesis}
\end{equation}
 if we consider the halo to no longer be perturbed after $2\tau_\text{dyn}$. In Fig.~\ref{fig:s_almm} we have delimited these two regions using black horizontal lines.
 
In the context of detection theory \citep[see e.g.][]{kay1998fundamentals}, one defines some test statistic, 
\begin{equation}
    \Gamma \underset{\mathcal{H}_0}{\overset{\mathcal{H}_1}{\gtrless}} \Gamma_\text{th},
\end{equation}
from the observed data such that when compared to a threshold, $\Gamma_\text{th}$, allows us to distinguish between the two hypotheses. 

In the following we will explore multiple ways of defining the test statistic and associated thresholds. This may appear cumbersome, however it is necessary to unambiguously define thresholds according to probabilistic criteria rather than arbitrary ones, while the variety of approaches we adopt allow us to check their robustness. 
 
\subsection{Frequentist Approach}
\label{sec:frequentist}
\begin{figure}
    \centering
    \includegraphics[width = 0.9\linewidth]{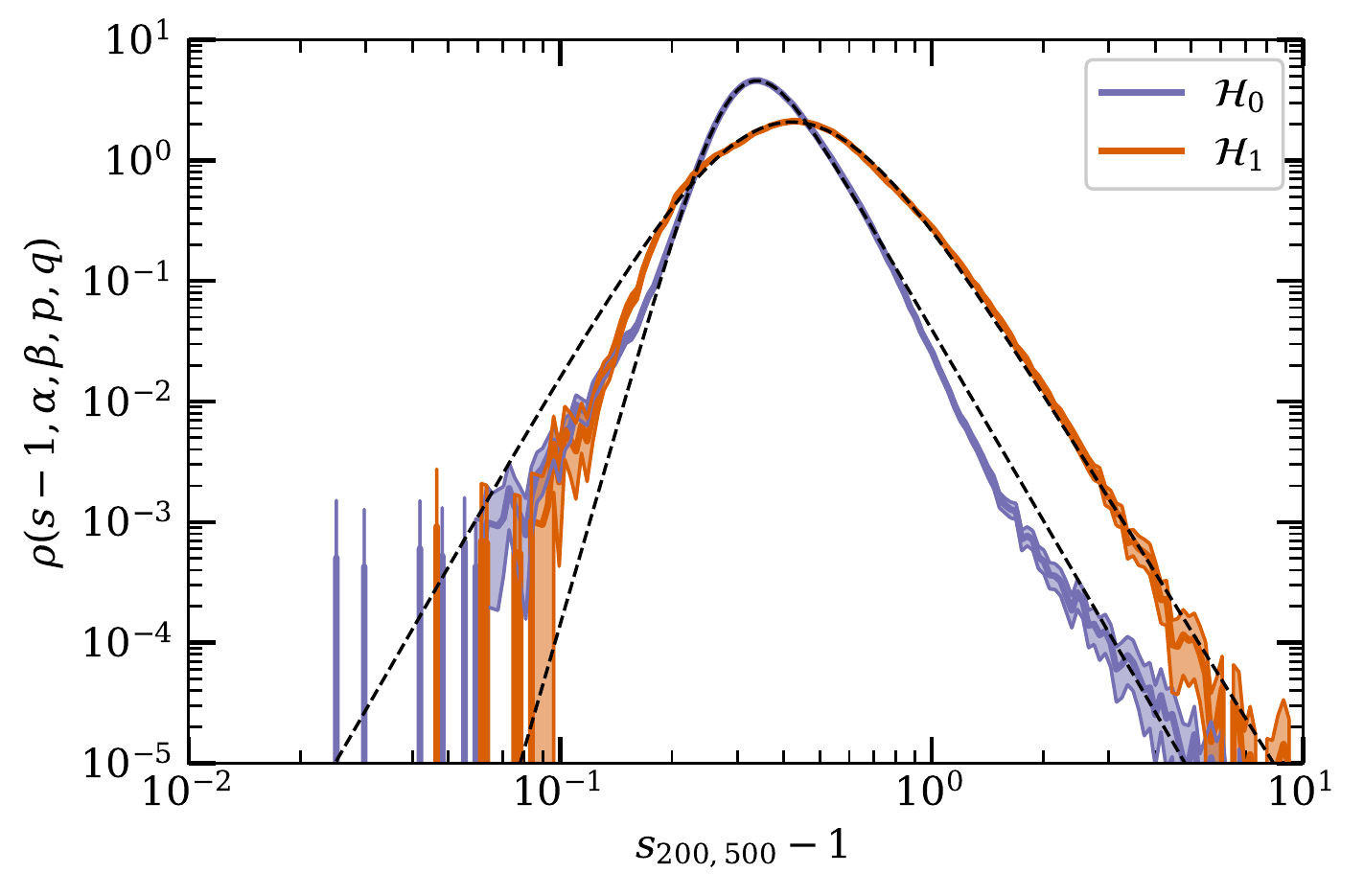}
    \caption{Estimated probability distribution functions for $\mathcal{H}_0$ (purple solid line) and $\mathcal{H}_1$ (orange solid line) hypotheses at $z=0$ along with best fitting generalised beta prime distribution functions (dotted black lines). The shaded area corresponds to the 95 percent confidence interval estimated over 1000 bootstrap iterations.}
    \label{fig:pdf_fit}
\end{figure}

We start with the simplest possible choice that is using $s_{200,500}$ as our test statistic. Separating our data set into the realisations of the two hypotheses, we estimate their respective likelihood functions, which we model using a generalised $\beta '$ probability density function (pdf),
\begin{equation}
    \rho(x,\alpha,\beta,p,q) = \frac{p\left(\frac{x}{q}\right)^{\alpha p - 1}\left(1+\left(\frac{x}{q}\right)^p\right)^{-\alpha-\beta}}{q\,B(\alpha,\beta)},
    \label{eq:gen_beta_prime}
\end{equation}
where $B(\alpha,\beta)$ is the Beta function and where $x = s_{200,500} - 1$. From our two samples we then fit this model using a standard least squares method to obtain the set of best fitting parameters under both hypotheses, these are reproduced in Tab.~\ref{tab:fit_params} and particular fits are shown in Fig.~\ref{fig:pdf_fit} for the halo catalogues at $z=0$. In both cases we additionally report the 95 percent confidence intervals estimated using 1000 bootstrap iterations. 

\begin{table}
    \centering
    \caption{Best fitting parameters for the distribution of sparsities at $z=0$ under both hypotheses. Here, we quote each parameter with its 95 percent confidence interval estimated over 1000 bootstrap iterations.}
    \begin{tabular}{r|cc}
    \hline Parameter & $\mathcal{H}_0$& $\mathcal{H}_1$\\ 
    \hline $\alpha$ & $1.4^{+0.1}_{-0.1}$ & $1.5^{+0.2}_{-0.2}$ \\
         $\beta$ & $0.61^{+0.03}_{-0.03}$ & $0.71^{+0.10}_{-0.08}$ \\
         $p$ & $7.7^{+0.3}_{-0.3}$ & $4.1^{+0.4}_{-0.3}$ \\
         $q$ & $0.304^{+0.002}_{-0.003}$ & $0.370^{+0.008}_{-0.008}$ \\
    \hline
    \end{tabular}
    \label{tab:fit_params}
\end{table}

\begin{figure}
    \centering
    \includegraphics[width = 0.9\linewidth]{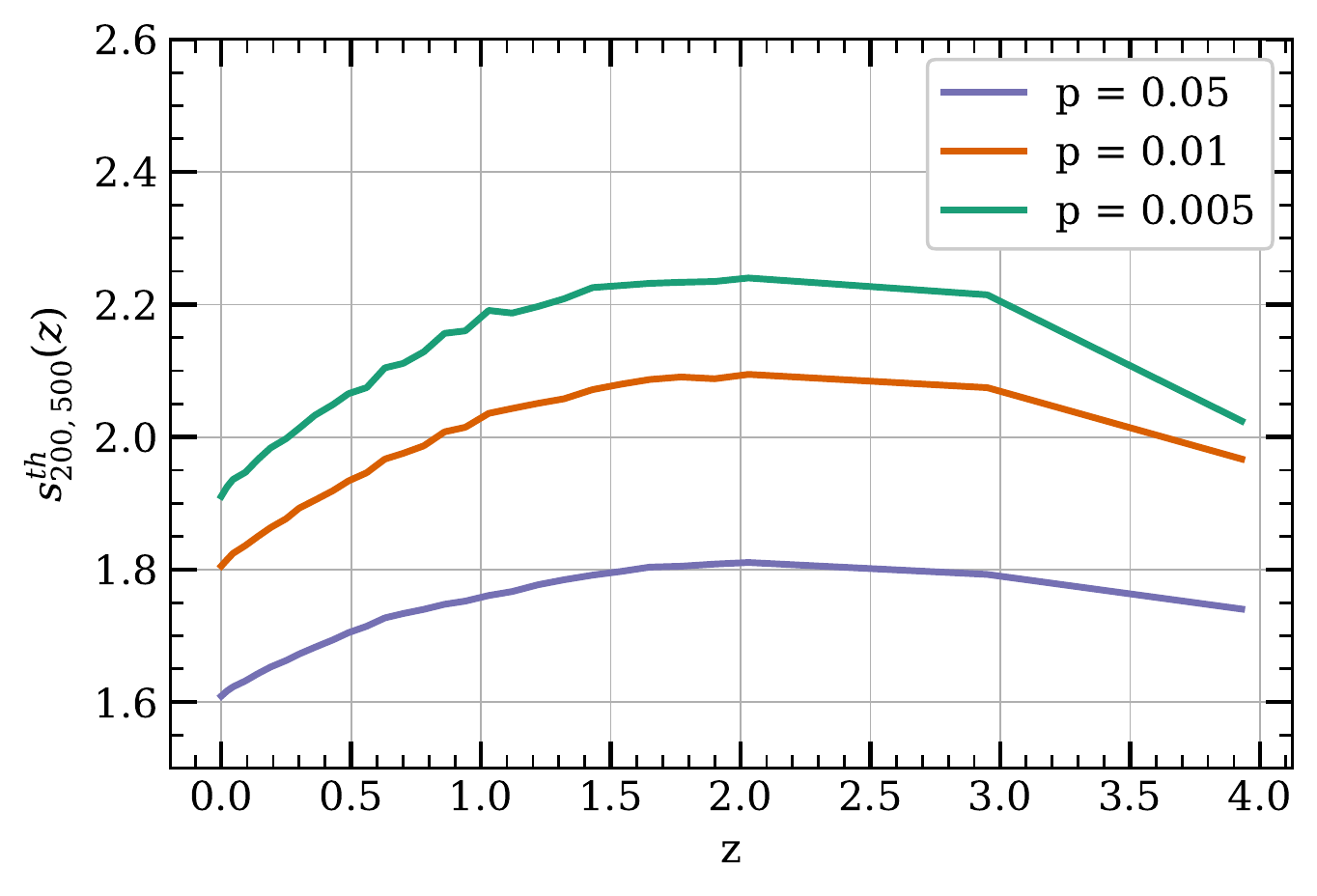}
    \caption{Sparsity thresholds $s^{\rm th}_{200,500}$ as function of redshift for ${\rm p}-$values$=0.05$ (purple solid line), $0.01$ (orange solid line) and $0.005$ (green solid line) computed using the Frequentist Likelihood-Ratio approach.}
    \label{fig:xi_of_z}
\end{figure}

The quality of the fits degrades towards the tails of the distributions, most notably under $\mathcal{H}_1$ due to the fact we do not account for the pulse feature. Nonetheless, they still allow us to obtain an estimate,  $\tilde\Sigma(x)$, of the corresponding likelihood ratio (LR) test statistic $\Sigma(x) = {\rho(x|\mathcal{H}_1)}/{\rho(x|\mathcal{H}_0)}$. Under the Neyman-Pearson lemma \citep[see e.g.][]{kay1998fundamentals} the true LR test statistic constitutes the most powerful estimator for a given binary test. We can express this statistic in terms of the fitted distribution, for $z=0$ this reads as:
\begin{align}
    \tilde\Sigma(x) &\propto x^{\alpha_1 p_1 - \alpha_0 p_0}\frac{(1 + (x/q_1)^{p_1})^{-\alpha_1 - \beta_1}}{(1 + (x/q_0)^{p_0})^{-\alpha_0 - \beta_0}}\\
    &= x^{-4.6}\frac{(1 + (x/0.370)^{4.1})^{-2.2}}{(1 + (x/0.304)^{7.7})^{-2.0}}
\end{align}
from which we can obtain an approximate expression, $\tilde\Sigma(x) \propto x^{1.8}$, for large values, $x \gg 0.3$. What one can observe is that for large values of sparsity the LR test statistic is a monotonously increasing function of $x = s_{200,500} - 1$, indicating that in this regime the sparsity itself will have comparable differentiating power to the LR test. A similar dependence holds at $z>0$. What this indicates is that we can use $\Gamma = s_{200,500}$ to efficiently differentiate between haloes that have undergone a recent major merger from an at rest population. In addition to this result, one can estimate a simple ${\rm p}-$value,
\begin{equation}
    {\rm p} = \text{P}_\text{r}(\Gamma > s_{200,500}|\mathcal{H}_0) = 1 - \int_0^{s_{200,500}-1}\rho(x|\mathcal{H}_0)dx
\end{equation}
i.e. the probability of finding a higher value of $s_{200,500}$ in a halo at equilibrium. And conversely, one can also estimate the value of the threshold corresponding to a given $p-$value, by inverting this relation. 
In Fig.~\ref{fig:xi_of_z} we have estimated the threshold corresponding to three key p-values at increasingly higher redshifts.
Here, each point is estimated using the sparsity distributions from the numerical halo catalogues. This figure allows to quickly estimate the values of sparsity above which a halo at some redshift $z$ should be considered as recently perturbed. 

It is worth noticing that these thresholds are derived from sparsity estimates from N-body halo masses. In contrast, sparsities of observed galaxy clusters are obtained from mass measurements that may be affected by systematic uncertainties that may differ depending on the type of observations. The impact of mass biases is reduced in the mass ratio, but it could still be present. As an example, using results from hydro/N-body simulations for an extreme model of AGN feedback model, \citet{Corasaniti2018} have shown that baryonic processes on average can bias the estimates of the sparsity $s_{200,500}$ up to $\lesssim 4\%$ and $s_{200,2500}$ up to $\lesssim 15\%$ at the low mass end. This being said, as long as the mass estimator is unbiased we expect our analysis to hold, albeit with a modification to the fitting parameters. In Section~\ref{testcase} we present a preliminary analysis of the impact of mass biasses on our approach, however we will leave more in depth investigations into this topic, as well as modifications that could arise from non-gravitational physics, to upcoming work.

\subsection{Bayesian approach}
\label{sec:Bayesisan}
An alternate way of tackling this problem is through the Bayesian flavour of detection theory. In this case, instead of looking directly at how likely the data $\boldsymbol{x}$ is described by a model characterised by the model parameters $\boldsymbol{\theta}$ in terms of the likelihood function $p(\boldsymbol{x}|\boldsymbol{\theta})$, one is interested in how likely is the model given the observed data, that is the posterior function $p(\boldsymbol{\theta}|\boldsymbol{x})$. 

Bayes theorem allows us to relate these two quantities:
\begin{equation}
    p(\bmath{\theta}|x) = \frac{p(x|\bmath{\theta})\pi(\bmath{\theta})}{\pi(x)},
    \label{eq:posterior}
\end{equation}
where $\pi(\bmath{\theta})$ is the prior distribution for the parameter vector $\bmath{\theta}$ and 
\begin{equation}
    \pi(x) = \int p(x|\bmath{\theta})\pi(\bmath{\theta}) d\bmath{\theta},
\end{equation}
is a normalisation factor, known as evidence.

While this opens up the possibility of estimating the parameter vector, which we will discuss in sub-section~\ref{statmergerepoch}, this approach also allows one to systematically define a test statistic known as the Bayes Factor,
\begin{equation}
    B_\text{f} = \frac{\int_{V_1} p(\bmath{x}|\bmath{\theta})\pi(\bmath{\theta})d\bmath{\theta}}{\int_{V_0} p(\bmath{x}|\bmath{\theta})\pi(\bmath{\theta})d\bmath{\theta}},
\end{equation}
associated to the binary test. Here, we have denoted $V_1$ and $V_0$ the volumes of the parameter space respectively attributed to hypothesis $\mathcal{H}_1$ and $\mathcal{H}_0$.

In practice, to evaluate this statistic we first need to model the likelihood. Again we use the numerical halo catalogues as calibrators. We find that the distribution of $s_{200,500}$ for a given value of the scale factor at the epoch of the last major merger, $a_\text{LMM}$, is well described by a generalised $\beta '$ pdf. In particular, we fit the set of parameters $\bmath{\theta} = [\alpha, \beta, p, q]^\top$ that depend solely on $a_\text{LMM}$ by sampling the posterior distribution using Monte-Carlo Markov Chains (MCMC) with a uniform prior for $a_\text{LMM}\sim \mathcal{U}(0; a(z))$\footnote{The upper bound is the scale factor at the epoch at which the halo is observed.}. This is done using the \textsc{emcee}\footnote{\href{https://emcee.readthedocs.io/en/stable/}{https://emcee.readthedocs.io/en/stable/}} library \citep{Emcee2013}. The resulting values of $B_\text{f}$ can then be treated in exactly the same fashion as the Frequentist statistic. It is however important to note that the Bayes factor is often associated with a standard ``rule of thumb'' interpretation \citep[see e.g.][]{Trotta2007} making these statistic particularly interesting to handle.

One way of comparing the efficiency of different tests is to draw their respective Receiver Operating Characteristic (ROC) curves \citep{Fawcett2006}, which show the probability of having a true detection, $\text{P}_\text{r}(\Gamma > \Gamma_{\rm th}|\mathcal{H}_1)$, plotted against the probability of a false one,  $\text{P}_\text{r}(\Gamma > \Gamma_{\rm th}|\mathcal{H}_0)$ for the same threshold. In other words, this means we are simply plotting the probability of finding a value of $\Gamma$ that is larger than the threshold under the alternate hypothesis against that of finding a value of $\Gamma$ larger than the same threshold under the null hypothesis. The simplest graphical interpretation of this type of figure is, the closer a curve gets to the top left corner the more powerful the test is at differentiating between both cases. 

\begin{figure}
    \centering
    \includegraphics[width = 0.9\linewidth]{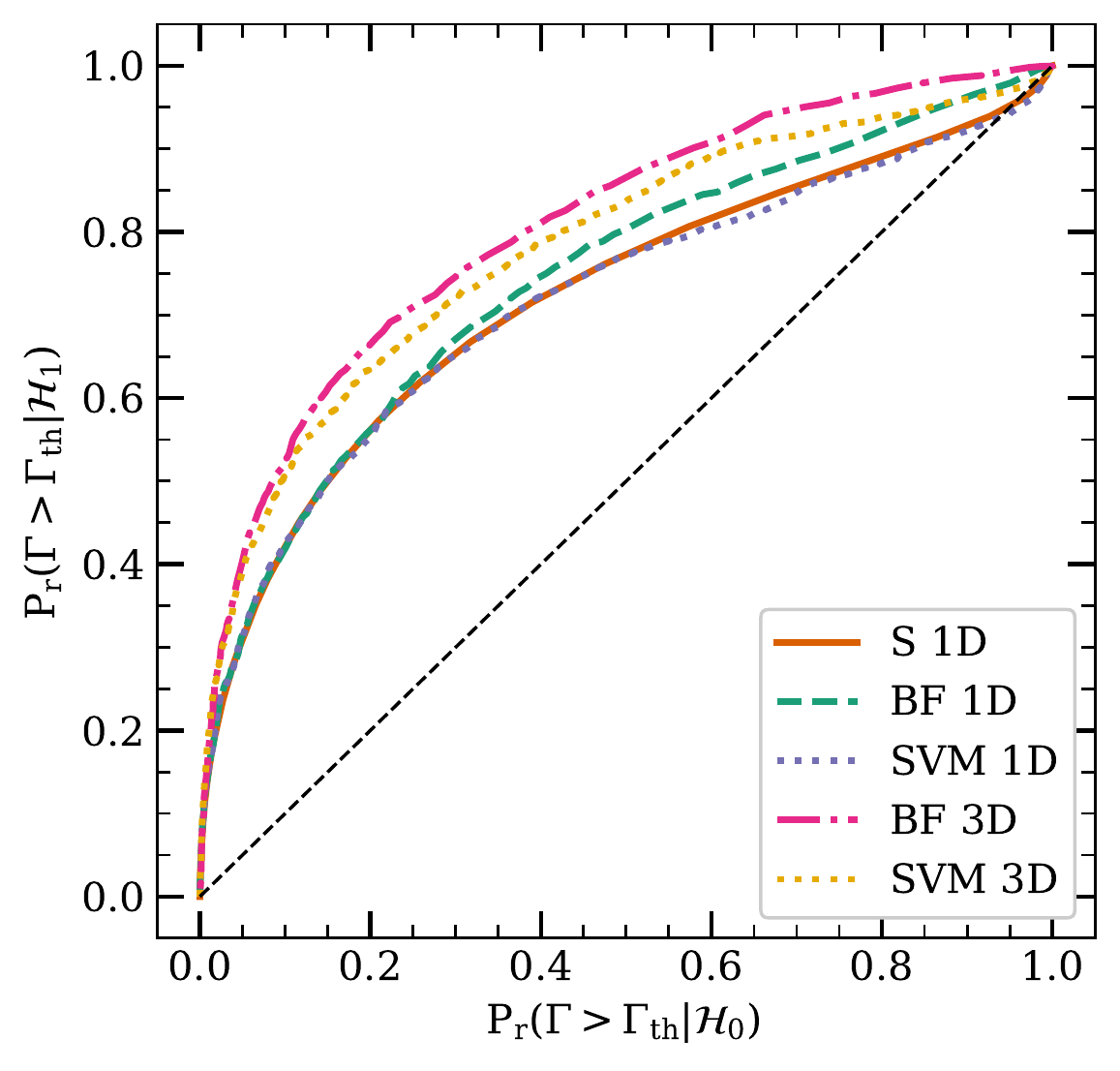}
    \caption{ROC curves associated with the binary tests studied in this work: the Frequentist sparsity test (S 1D, solid orange line), the Bayes Factor based on a single sparsity value (BF 1D, dashed green line) and using three values (BF 3D, dash-dotted magenta line), the Support Vector Machines with one sparsity value (SVM 1D, dotted purple line) and three sparsities (SVM 3D, dotted yellow line). What can be observed is that all 1D tests are equivalent at small false alarm rates and the only way to significantly increase the power of the test is to increase the amount of input data, i.e. adding a third mass measurement as in the BF 3D and SVM 3D cases.}
    \label{fig:roc_curves}
\end{figure}

In Fig.~\ref{fig:roc_curves} we plot the ROC curves corresponding to all the tests we have studied in the context of this work. These curves have been evaluated using a sub sample of $10^4$ randomly selected haloes from the MDPL2 catalogues at $z=0$ with masses $M_{200\text{c}} > 10^{13}\,h^{-1}\text{M}_{\odot}$. Let us focus on the comparison between the Frequentist direct sparsity approach (S 1D) against the Bayes Factor obtained using a single sparsity measurement (BF 1D). We can see that both tests have very similar ROC curves for low false alarm rates. This indicates that we do not gain any substantial power from the additional computational work done to estimate the Bayes factor using a single value of sparsity.

While this may seem as the end of the line for the method based on the Bayes factor, the latter does present the significant advantage of being easily expanded to include additional data. In our case this comes in the form of adding additional sparsity measurements at different overdensities. Simply including a third mass measurement, here we use $M_{2500\text{c}}$, gives us access to two additional sparsities from the three possible pairs, $s_{200,500},\,s_{200,2500}$ and $s_{500,2500}$. This leads us to defining each halo as a point in a 3-dimensional space with coordinates
\begin{equation}
    \begin{cases}
    x = s_{200,500} - 1 \\
    y = s_{200,2500} -1 \\
    z = s_{500,2500} -1
    \end{cases}
\end{equation}
After estimating the likelihood in this coordinate system, one quickly observes that a switching spherical-like coordinate system, $\mathbfit{r} = [r, \vartheta, \varphi]^\top$,  allows for a much simpler description. The resulting likelihood model,
\begin{equation}
    L(\mathbfit{r};\bmath{\theta},\bmath{\mu},\mathbfss{C}) = \frac{f(r;\bmath{\theta})}{2\pi\sqrt{|\mathbfss{C}|}}\exp\left[-\frac{1}{2}(\bmath{\alpha} - \bmath{\mu})^\top\mathbfss{C}^{-1}(\bmath{\alpha} - \bmath{\mu})\right],
    \label{eq:like3D}
\end{equation}
treats $r$ as independent from the two angular coordinates that are placed within the 2-vector $\bmath{\alpha} = [\vartheta, \varphi]^\top$. Making the radial coordinate independent allows us to constrain $f(r,\bmath{\theta})$ simply from the marginalised distribution. Doing so we found that the latter is best described by a Burr type XII \citep{10.1214/aoms/1177731607} distribution,
\begin{equation}
    f(x,c,k,\lambda,\sigma) = \frac{ck}{\sigma}\left(\frac{x-\lambda}{\sigma}\right)^{c-1}\left[1+\left(\frac{x-\lambda}{\sigma}\right)^2\right]^{-k-1},
\end{equation}
with additional displacement, $\lambda$, and scale, $\sigma$, parameters. In total the likelihood function is described by 9 parameters, 3 of which are constrained by fitting the marginalised distribution of $r$ realisations assuming $\lambda = 0$ and 5, 2 in $\bmath{\mu}$ and 3 in $\mathbfss{C}$, are measured through unbiased sample means and covariances. 

In a similar fashion as in the single sparsity case, we evaluate these parameters as functions of $a_\text{LMM}$ and thus recover a posterior likelihood for the epoch of the last major merger using MCMC, again applying a flat prior on $a_\text{LMM}$. This posterior in turn allow us to measure the the corresponding Bayes Factor. We calculate these Bayes factors for the same test sample used previously and evaluate the corresponding ROC curve (BF 3D in Fig.~\ref{fig:roc_curves}). As intended, the additional mass measurement has for effect of increasing the detection power of the test, thus raising the ROC curve with respect to the 1D tests. Increasing the true detection rate from 40 to 50 percent for a false positive rate of 10 percent. We have tested that the same trends hold valid at $z>0$.

\subsection{Support Vector Machines}
An alternative to the Frequentist -- Bayesian duo is to use machine learning techniques designed for classification. While Convolutional Neural Networks \citep[see eg.][for a review]{2015Natur.521..436L} are very efficient and have been profusely used to classify large datasets, both in terms of dimensionality and size, recent examples in extra-galactic astronomy include galaxy morphology classification \citep[e.g.][]{Hocking2018,Martin2020,Abul_Hayat2020,Cheng2021,Spindler2021} detection of strong gravitational lenses \citep[e.g.][]{Jacobs2017,Jacobs2019, Lanusse2018,Canameras2020,Huang2020,Huang2021,He2020,Gentile2021,Stein2021} galaxy merger detection \citep{Ciprijanovic2021} and galaxy cluster merger time estimation \citep{Koppula2021}. They may not be the tool of choice when dealing with datasets of small dimensionality, like the case at a hand. A simpler option for this problem is to use Support Vector Machines (SVM) \citep[see e.g.][]{Cristianini2000} as classifiers for the hypotheses defined in Eq.~(\ref{eq:hypothesis}), using as training data the sparsities measured from the halo catalogues.

A SVM works on the simple principle of finding the boundary that best separates the two hypotheses. In opposition to Random Forests \citep[see e.g.][]{Breiman2001} which can only define a set of horizontal and vertical boundaries, albeit to arbitrary complexity, the SVM maps the data points to a new euclidean space and solves for the plane best separating the two sub-classes. This definition of a new euclidean space allows for a non-linear boundary between the classes. For large datasets however the optimisation of the non-linear transformation can be slow to converge, and thus we restrict ourselves to a linear transformations. To do so we make use the \textsc{scikit-learn}\footnote{\href{https://scikit-learn.org/}{https://scikit-learn.org/stable/}} \citep{scikit-learn} python package. The ``user friendly'' design of this package allows for fast implementations with little required knowledge of python and little input from the user, giving this method an advantage over its Frequentist and Bayesian counterparts.

In order to compare the effectiveness of the SVM tests, with 1 and 3 sparsities, against those previously presented we again plot the corresponding ROC curves\footnote{Note that the test data used for the ROC curves was excluded from the training set.} in Fig.~\ref{fig:roc_curves}. What can be seen is that the SVM tests reach comparable differentiating power to both the Bayesian and Frequentist test for 1 sparisty and is only slightly out performed by the Bayesian test using 3 sparsities. This shows that designing a statistical test based on the sparsity can be done in a simple fashion without significant loss of differentiation power. Making sparsity an all the more viable proxy to identify recent major mergers.

\subsection{Estimating cluster major merger epoch}\label{statmergerepoch}

In the previous section we have investigated the possibility of using halo sparsity as a statistic to identify clusters that have had a recent major merger. We will now expand the Bayesian formulation of the binary test to \emph{estimate} when this last major merger took place. This can be achieved by using the posterior distributions which we have previously computed to calculate the Bayes Factor statistics. These distributions allow us to define the most likely epoch for the last major merger as well as the credible interval around this epoch. 

\begin{figure}
    \centering
    \includegraphics[width = 0.95\linewidth]{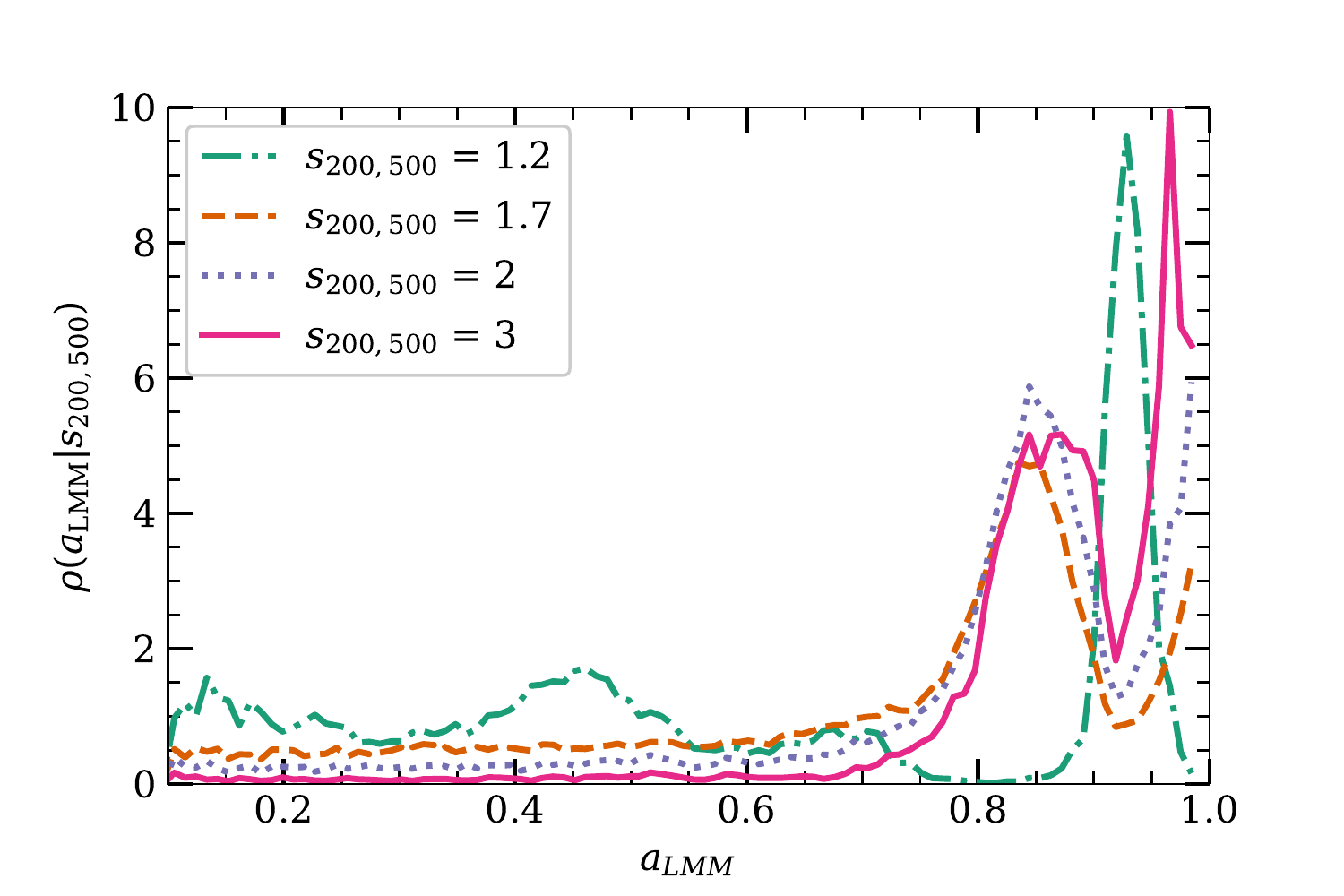}
    \caption{Posterior distributions for different values of the sparsity $s_{200,500}=1.2$ (dash-dotted green line), $1.7$ (dashed orange line), $2$ (dotted purple line) and $3$ (solid magenta line). We can see that for large sparsity values, the distributions are bimodal at recent epoch, while low values produce both a continuous distribution at low scale factor values as well as a single peak at recent epochs corresponding to a confusion region. This induce a degeneracy that needs to be broken if we are to accurately estimate $a_\text{LMM}$.}
    \label{fig:post_1sparsity}
\end{figure}

Beginning with the one sparsity estimate, in Fig.~\ref{fig:post_1sparsity} we plot the resulting posterior distributions $p(a_{\rm LMM}|s_{200,500})$ obtained assuming four different values of $s_{200,500}=1.2,1.7,2$ and $3$ at $z=0$. As we can see, in the case of large sparsity values ($s_{200,500}\ge 1.7$), we find a bimodal distribution in the posterior, caused by the pulse-like feature in the structure of the joint distribution shown in Fig.~\ref{fig:sva} and which is consequence of universal imprint of the major merger on the halo sparsity evolution shown in Fig.~\ref{fig:sparsity_histories_2}. In particular, we notice that the higher the measured sparsity and the lower the likelihood of having the last major merger to occur in the distant past. A consequence of this pulse-like feature is that a considerable population of haloes with a recent major mergers, characterised by $-1/2 <T(a_\text{LMM}; a(z))<-1/4$, have sparsities in the same range as those in the quiescent regime. This confusion region results in a peak of the posterior distribution for the $s_{200,500}=1.2$ case, that is located on the minimum of the binomial distributions associated to the values of $s_{200,500}\ge 1.7$. This suggests the presence of a degeneracy, such that quiescent haloes may be erroneously identified as haloes having undergone a recent merger, or on the contrary haloes having undergone a recent merger are misidentified as quiescent haloes.

\begin{figure*}
    \centering
    \includegraphics[width = 0.95\linewidth]{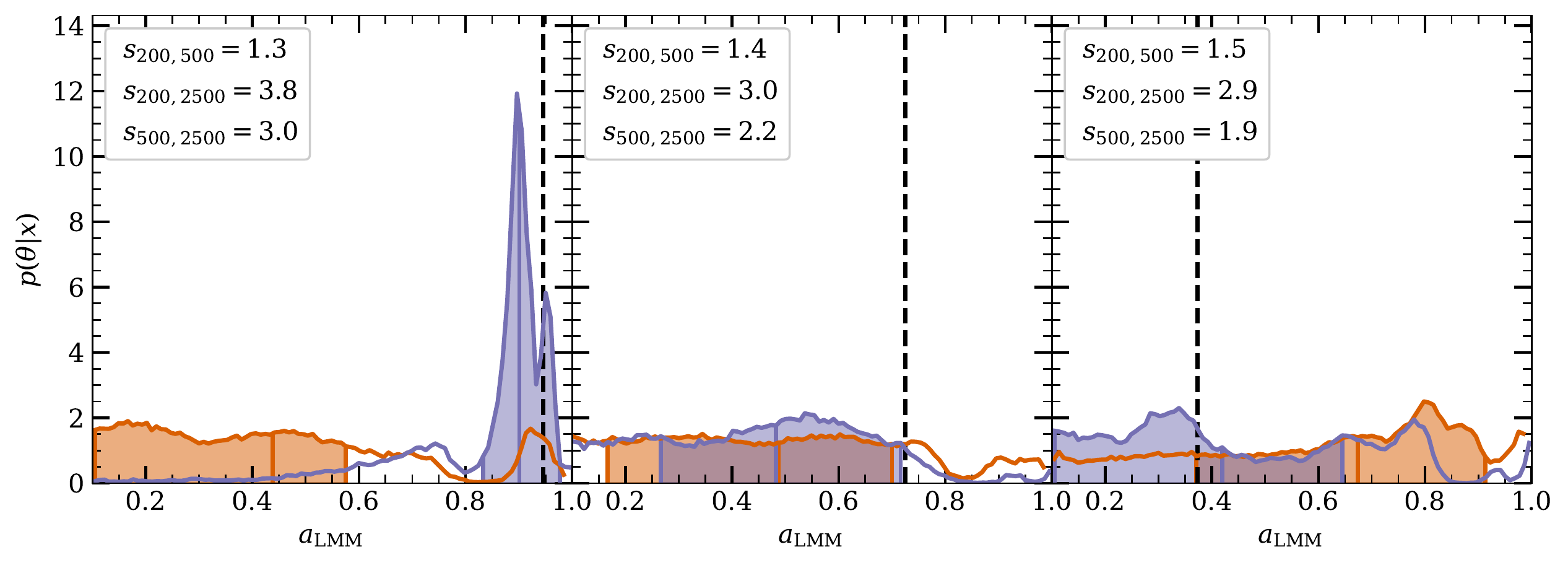}
    \caption{Posterior distributions of the last major merger epoch for three selected haloes with different sparsity values from the $z=0$ halo catalogue. The shaded areas corresponds to the 68\% credible interval around the median (coloured vertical line) assuming a single (orange) and three sparsity (purple) measurements. The black vertical dashed lines mark the true fiducial value of $a_\text{LMM}$ for each of the selected haloes.}
    \label{fig:post_3sparsity}
\end{figure*}

The presence of this peak in the $p(a_{\rm LMM}|s_{200,500})$ posterior for low sparsity values biases the Bayesian estimation towards more recent major mergers when using a single sparsity measurement. As a result the previously mentioned Bayes factors, which depend on such a posterior, will also be biased towards recent mergers resulting in higher measured values. Moreover, this impacts our choice when it comes to the estimation we use, indeed a maximum likelihood estimation will be very sensitive to this peak. Therefore, we prefer to use a median likelihood estimation that is significantly more robust. The credible interval is then estimated iteratively around the median as to encompass 68 percent of the total probability. The end result of this procedure is shown in Fig.~\ref{fig:post_3sparsity} where we plot inferred posteriors along with the corresponding credible intervals (shaded areas) and median likelihood measurements (vertical lines) obtained assuming one (orange curves) and three sparsity (purple curves) values from three haloes selected from the numerical halo catalogue at $z=0$. The black vertical dashed lines indicate the true $a_{\rm LMM}$ value of the haloes.
We can clearly see that the inclusion of an additional mass measurement (or equivalently two additional sparsity estimates) allows to break the $s_{200,500}$ degeneracy between quiescent and merging haloes with low sparsity values. In such a case the true $a_{\rm LMM}$ value is found to be within the $1\sigma$ credible regions. Hence, this enable us to also identify merging haloes that are located in the confusion region.

\section{Cosmological Implications}\label{cosmo_imp}

Before discussing practical applications on the use of large halo sparsities as tracers of major merger events in clusters, it is worth highlighting the impact that such systems can have on average sparsity measurements that are used for cosmological parameter inference. 

\begin{figure}
    \centering
    \includegraphics[width = \linewidth]{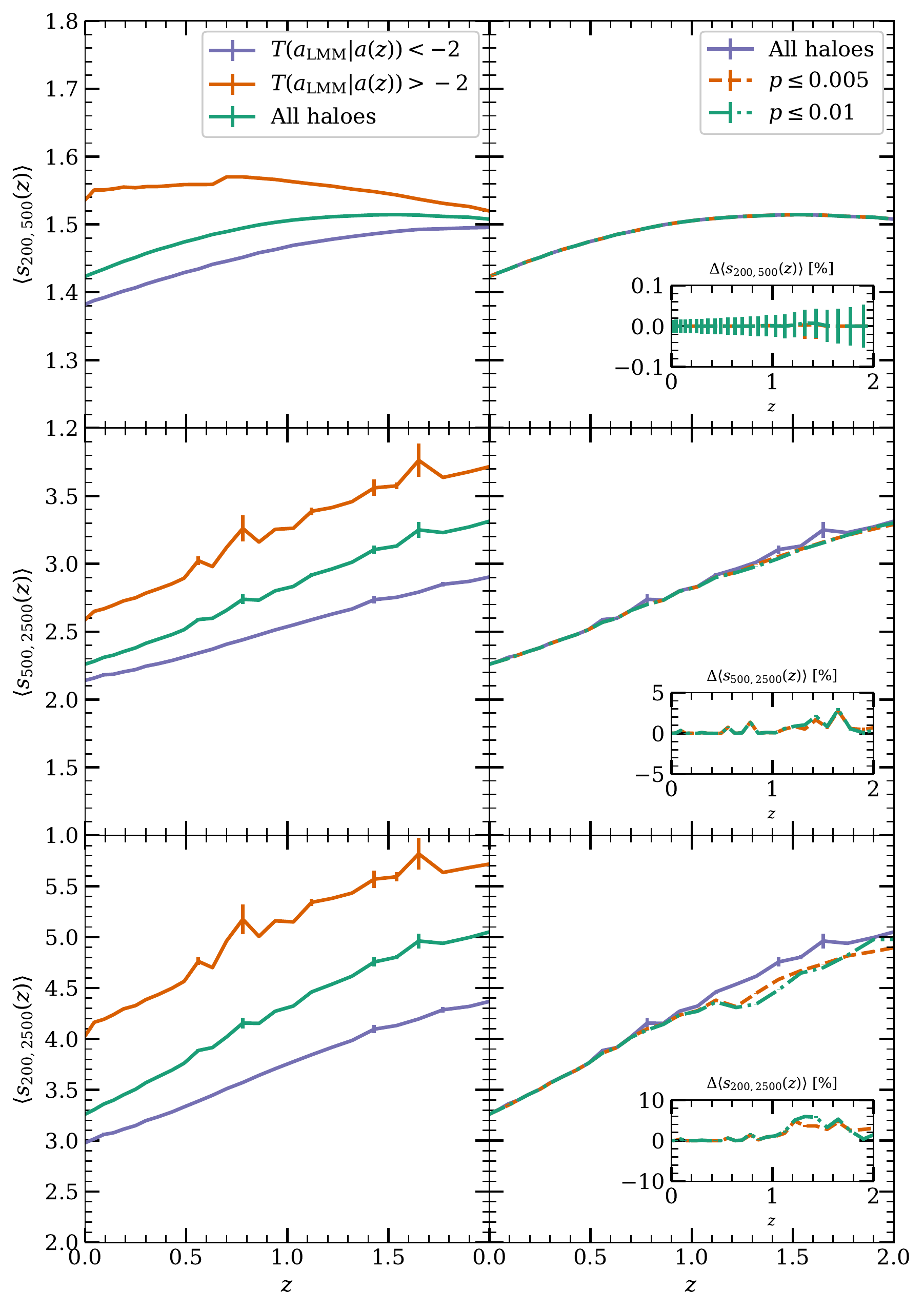}
    \caption{Redshift evolution of the average halo sparsity $\langle s_{200,500}\rangle$ (top panels), $\langle s_{500,2500}\rangle$ (middle panels) and $\langle s_{200,2500}\rangle$ (bottom panels). In the left panels we show the average sparsity estimated for the full halo samples (green curves), for haloes which are within two dynamical times from the last major merger event (blue curves) and for haloes which are at more than two dynamical times from it (orange curves). In the right panels we show the average sparsity estimate from the full halo samples (blue curves) and for selected samples from which we removed outliers whose sparsity lies above thresholds corresponding to p-values of $p\le 0.01$ (green curves) and $p\le 0.005$ (orange curves). In the inset plots we show the relative differences with respect to the mean sparsity estimated from the full catalogues.}
    \label{fig:cosmo_mean}
\end{figure}

Halo sparsity depends on the underlying cosmological model \citep{Balmes2014,Ragagnin2021}, and it has been shown \citep[][]{Corasaniti2018,Corasaniti2021} that the determination of the average sparsity of an ensemble of galaxy clusters estimated at different redshifts can provide cosmological constraints complementary to those from standard probes. This is possible thanks to an integral relation between the average halo sparsity at a given redshift and the halo mass function at the overdensities of interest, which allows to predict the average sparsity for a given cosmological model \citep{Balmes2014}. Hence, the average is computed over the entire ensemble of haloes as accounted for by the mass functions. In principle, this implies that at a given redshift the mean sparsity should be computed over the available cluster sample without regard to their state, since any selection might bias the evaluation of the mean. This can be seen in the left panel of Fig.~\ref{fig:cosmo_mean}, where we plot the average sparsity $\langle s_{200,500}\rangle$ (top panel), $\langle s_{500,2500}\rangle$ (central panel) and $\langle s_{200,2500} \rangle$ as function of redshift in the case of haloes which are within two dynamical times since the last major merger (blue curves), for those which are more than two dynamical times since the last major merger (orange curves) and for the full sample (green curves). As we can see removing the merging haloes induces a $\sim 10\%$ bias on $\langle s_{200,500}\rangle$ at $z=0$, which decreases to $\sim 4\%$ at $z=1$, while in the same redshift range the bias is at $\sim 20\%$ level for $\langle s_{500,2500}\rangle$ and $\sim 30\%$ for $\langle s_{200,2500}\rangle$. 

However, the dynamical time is not observable and in a realistic situation, one might have to face the reverse problem, which is that of having a number of outliers characterised by large sparsity values in a small cluster sample, potentially biasing the estimation of the mean compared to that of a representative cluster ensemble. Which clusters should be considered as outliers, and which should be removed from the cluster sample such that the estimation of the mean sparsity will remain representative of the halo ensemble average, say at sub-percent level? To assess this question, we can make use of the sparsity thresholds defined in Section~\ref{sec:frequentist} based on the p-value statistics. As an example in the right panel of Fig.~\ref{fig:cosmo_mean}, we plot the mean sparsities $\langle s_{200,500}\rangle$, $\langle s_{500,2500}\rangle$ and $\langle s_{200,2500}\rangle$ as function of redshift computed using the full halo sample (blue curves), and a selected halo sample from which we have removed haloes with sparsities above the sparsity thresholds, such as those shown in Fig.~\ref{fig:xi_of_z}, associated to p-values of $p\le 0.01$ (green curves) and $p\le 0.005$ (orange curves) respectively. We can see that removing outliers alter the estimated mean sparsity $\langle s_{200,500}\rangle$ at sub-percent level over the range $0<z<2$, and in the case of $\langle s_{500,2500}\rangle$ and $\langle s_{200,2500}\rangle$ up to a few per-cent level only in the high-redshift range $1\lesssim z <2$.

\section{Towards practical applications}\label{testcase}

We will now work towards applying the statistical analysis presented in Section~\ref{calistat} to observational data. To this purpose we have specifically developed the numerical code \textsc{lammas}\footnote{\href{https://gitlab.obspm.fr/trichardson/lammas}{https://gitlab.obspm.fr/trichardson/lammas}}. Given the mass measurements $M_{200\text{c}}$, $M_{500\text{c}}$ and $M_{2500\text{c}}$ of a galaxy cluster, the code computes the sparsity data vector  $\bmath{D}=\{s_{200,500},s_{200,2500},s_{500,2500}\}$ (the last two values only if the estimate of $M_{2500\text{c}}$ is available) and performs a computation of the frequentist statistics discussed in Section~\ref{sec:frequentist} and the Bayesian computation presented in Section~\ref{sec:Bayesisan}. The code computes the frequentist  p-value only for $s_{200,500}$ and it's associated uncertainty. Bayesian statistics are measure for both 1 and 3 sparsities, these include the posterior distributions $p(a_{\rm LMM}|\bmath{D})$ and their associated marginal statistics, along with the Bayes factor, $B_\text{f}$, using the available data. We implement the statistical distributions of merging and quiescent halo populations calibrated on the halo catalogues from the Uchuu simulations \citep{Ishiyama2021} rather than MDPL2, thus benefiting from the higher mass resolution and redshift coverage of the Uchuu halo catalogues. A description of the code \textsc{lammas} is given in Appendix~\ref{LAMMAS}. In the following  we will first validate this code by presenting it with haloes from N-body catalogues that were not used for calibration. We will then quantify the robustness of our analysis to observational mass biases using empirical models. In particular, we will focus on weak lensing, hydrostatic equilibrium and NFW-concentration derived galaxy cluster masses. Finally we present a preliminary analysis of two galaxy clusters, Abell 383 and Abell 2345.

\subsection{Validation on simulated haloes}

As we have calibrated \textsc{lammas} using the Uchuu simulation suite \citep{Ishiyama2021}, we use a randomly selected sample of $10^4$ haloes from the previously described MDPL2 catalogues as validation dataset. This choice has two main advantages, firstly it naturally guarantees the same haloes are not used in both the calibration and the validation; secondly, it allows to test the robustness of the method to small changes in cosmology as the Uchuu suite is run on the cosmology of \citet{2016A&A...594A..13P} compared to MDPL2 which is run using that of \citet{2014A&A...571A..16P}. Furthermore, we choose to do this validation at $z=0.248$ to ensure that our pipeline also performs well at redshifts $z\neq 0$.

\begin{figure}
    \centering
    \includegraphics[width = \linewidth]{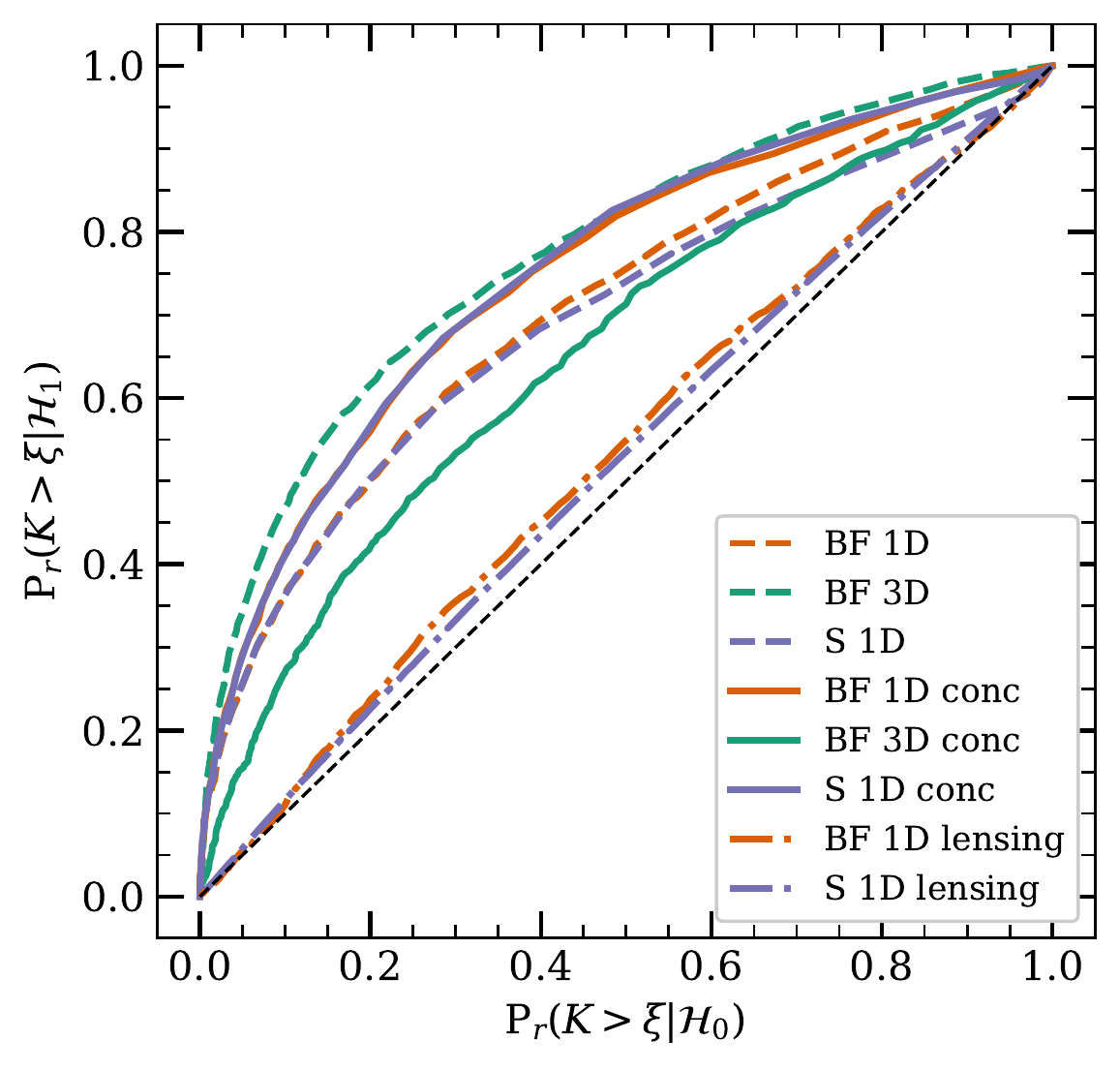}
    \caption{ROC curves estimated from the validation dataset for sparsities estimated from the N-body halo masses (dashed lines), from the concentration parameter of the best-fitting NFW-profile (solid lines) and in the case of a conservative model for the mass bias induced by lensing observations (dash-dotted lines) in the case of the single sparsity Bayesian (BF 1D, orange curves) and frequentist (S 1D, blue curves) estimators and the three sparsity Bayesian estimator (BF 3D, green curves).  We can see that in all cases, S 1D and BF 1D tests offer a similar detection power. Comparing the BF 3D curves to the S 1D ones, it is clear that while adding an independent sparsity measurement increases the detection power, this is not the case when the sparsities are deduced from the concentration parameter, with the latter having the opposite effect. Finally we can also see that strong mass biases have a strong negative impact on the efficiency of the detection of mergers.}
    \label{fig:validation_roc_curves}
\end{figure}

We evaluate the efficiency of the detection procedure in terms of ROC curves shown in Fig.~\ref{fig:validation_roc_curves} and constructed using the same method as those shown in Fig.~{\ref{fig:roc_curves}}. We plot the case of the single sparsity frequentist (S 1D) and Bayesian (BF 1D) estimators, as well as the three sparsity Bayesian (BF 3D) estimator for sparsity measurements inferred from N-body halo masses (dashed lines), lensing masses (dash-dotted lines) and NFW-concentration derived masses (solid lines). Comparing the dashed curves of Fig.~\ref{fig:validation_roc_curves} and those in Fig.~{\ref{fig:roc_curves}} we can see that for the validation dataset considered here the efficiency of merger detection of the different test statistics is comparable to that we have inferred for the MDPL2 halo sample.

\begin{figure}
    \centering
    \includegraphics[width = .9\linewidth]{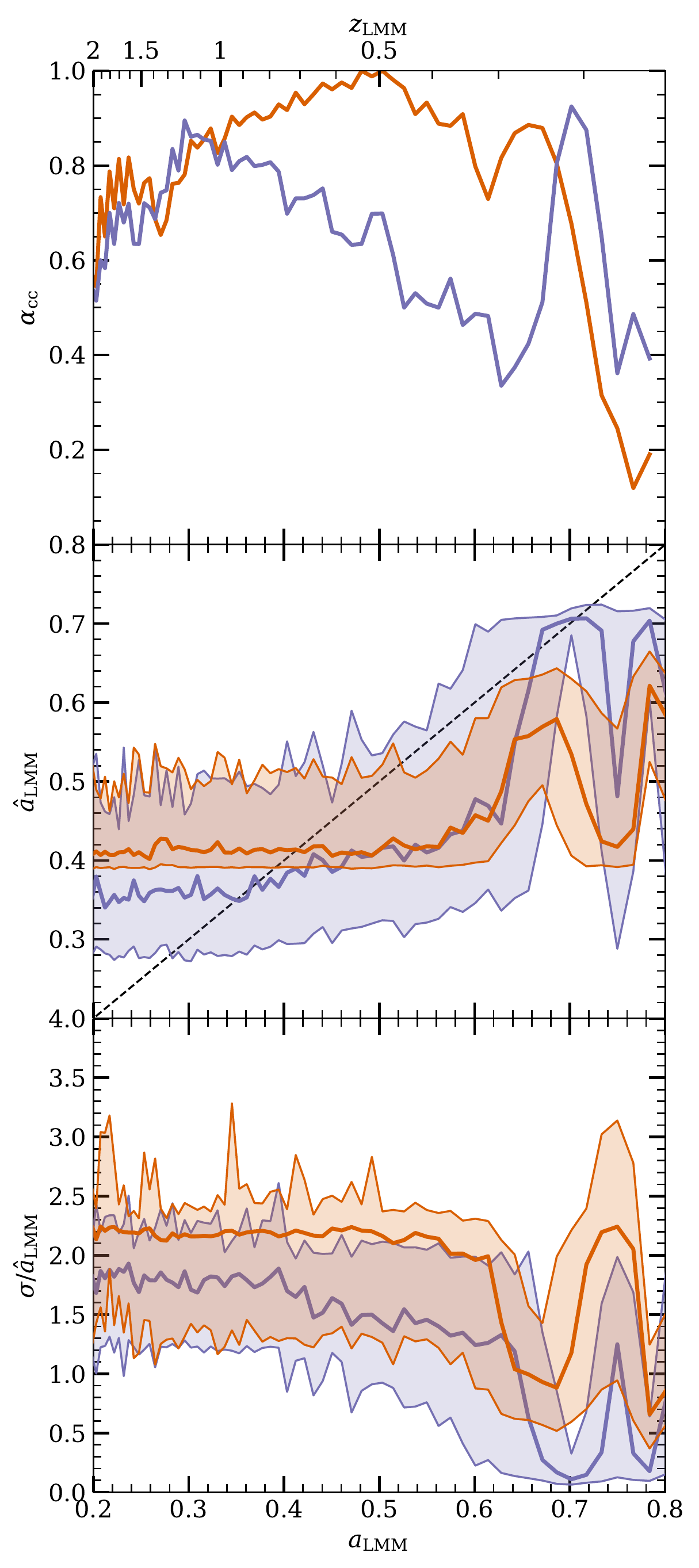}
    \caption{\textit{Top:} Accuracy of the estimation of the epoch of the last major merger, $\alpha_{\rm cc}$, as a function of the true value $a_{\rm LMM}$ of the haloes in the validation sample for both the 1S (orange solid line) and 3S (blue solid line) estimators respectively. \textit{Middle:} Median value of the estimated epoch of the last major merger, $\hat{a}_{\rm LMM}$, as function of the true value for the 1S (orange curves) and 3S (blue curves) estimators respectively.  The shaded areas correspond to the $68\%$ interval around the median, while the dashed diagonal line gives the ideal value of the estimator $\hat{a}_{\rm LMM}=a_{\rm LMM}$.  \textit{Bottom:} relative width of the $68\%$ interval around the median value of $\hat{a}_{\rm LMM}$ as a function of the true value $a_{\rm LMM}$ for the 1S (orange curves) and 3S (blue curves) estimators respectively. We refer the reader to the text for a detailed discussion of the various trends.}
    \label{fig:test_metrics}
\end{figure}

We quantify the accuracy of the estimation procedure by introducing three metrics defined as:
\begin{itemize}
    \item the accuracy as given by the frequency at which the true value $a_\text{LMM}$ of a halo is recovered within the $1\,\sigma$ credible interval, $\alpha_\text{cc}$;
    \item the estimated epoch of the last major merger, $\hat{a}_{\rm LMM}$;
    \item the relative width of the $1\,\sigma$ credible interval, $\sigma/\hat{a}_{\rm LMM}$.
\end{itemize}
In Fig.~\ref{fig:test_metrics}, we plot these metrics as function of the true scale factor (redshift) of the last major merger of the haloes in the validation sample for the case of a single sparsity (orange curves) and three sparsity (blue curves) measurements, to which we will simply refer as 1S and 3S respectively. At first glance, it may appear from the top panel as if the 1S estimator is more accurate at recovering the merger epoch than it's 3S counterpart over a large interval $0.2<a_{\rm LMM}<0.68$. However, this is simply due to the fact that for haloes which are more than two dynamical times from their last major merger the posterior distribution is nearly flat and the estimator returns the same estimated time, as can be seen from the plot in the central panel. Consequently, the increased accuracy is simply due to wider credible intervals, as can be seen in the bottom panel. Hence, in this particular regime it is more prudent to extract an upper bound on $\hat{a}_{\rm LMM}$ from the resulting posterior, rather than a credible interval.

We can see that the trend is reversed for recent mergers occurring at $0.68<a_{\rm LMM}>0.8$, with the 3S estimator being much more accurate at recovering the scale factor of the last major merger and with restricted error margins (see blue curves in top and bottom panels respectively). Nevertheless from the middle panel, we may notice that both the 1S and 3S estimators have an area of confusion around the dip of the pulse feature in the $\hat{a}_{\rm LMM}$ plot. In both cases, we see that the estimator disfavours very recent merger (at $a_{\rm LMM}\approx 0.8$ in favour of placing them in the second bump of pulse, thus causing the median value and the $68\%$ region of $\hat{a}_{\rm LMM}$ to be lower than the true value of the last major merger epoch. An effect, that should be kept in mind when using the pipeline.

\subsection{Systematic Bias}

The statistical methodology we have developed here relies on sparsity estimates from N-body halo masses. However, these masses are not directly comparable to those inferred from galaxy cluster mass measurements, since the latter involve systematic uncertainties that may bias the cluster mass estimates compared to that from dark matter only simulations. Hence, before applying the sparsity test to real observations, we check the robustness of our approach against observational mass biases. More specifically, we will review conservative estimates of these biases for various mass estimation techniques and attempt to quantify the effect that these have on the sparsity.

\subsubsection{Weak Lensing Mass Bias}
A well known source of systematic error in weak lensing mass estimates comes from fitting the observed tangential shear profile of a cluster with a spherically symmetric NFW inferred shear profile. In such a case deviations from sphericity of the mass distribution within the cluster, as well as projection effects induce a systematic error on the estimated cluster mass that may vary at different radii, consequently biasing the evaluation of the sparsity.

\citet{Becker2011} have investigated the impact of this effect on weak lensing estimated masses. They modelled the observed mass at overdensity $\Delta$ as:
\begin{equation}
    M_{\Delta}^{\text{WL}} = M_\Delta \exp(\beta_\Delta)\exp(\sigma_\Delta X),
\end{equation}
where $M_{\Delta}$ is the unbiased mass, $\beta_{\Delta}$ is a deterministic bias terms, while the third factor is a stochastic term with $\sigma_{\Delta}$ quantifying the spread of a log-normal distribution and $X\sim\mathcal{N}(0,1)$. Under the pessimistic assumption of independent scatter on both mass measurements, the resulting bias on the sparsity then reads as:
\begin{equation}\label{spars_wl_bias}
s_{\Delta_1,\Delta_2}^{\rm WL} = s_{\Delta_1,\Delta_2} \left(b^{\rm WL}_{\Delta_1,\Delta_2} +1\right) \exp\left(\sigma^{\rm WL}_{\Delta_1,\Delta_2} X\right), 
\end{equation}
where $b^{\rm WL}_{\Delta_1,\Delta_2} = \exp(\beta_{\Delta_1} - \beta_{\Delta_2}) - 1$ and $\sigma^{\rm WL}_{\Delta_1,\Delta_2} = \sqrt{\sigma_{\Delta_1}^2 + \sigma_{\Delta_2}^2}$, with the errors being propagated from the errors quoted on the mass biases. \citet{Becker2011} have estimated the mass bias model parameters at $\Delta_1=200$ and $\Delta_2=500$, using the values quoted in their Tab.~3 and 4 we compute the sparsity bias $b^{\rm WL}_{200,500}$ and the scatter $\sigma^{\rm WL}_{200,500}$, which we quote in Tab.~\ref{tab:WL_bias}, for different redshifts and galaxy number densities, $n_\text{gal}$, in units of galaxies per arcmin$^{-2}$. Notice that the original mass bias estimates have been obtained assuming an intrinsic shape noise $\sigma_e = 0.3$.

\begin{figure}
    \centering
    \includegraphics[width = 0.9\linewidth]{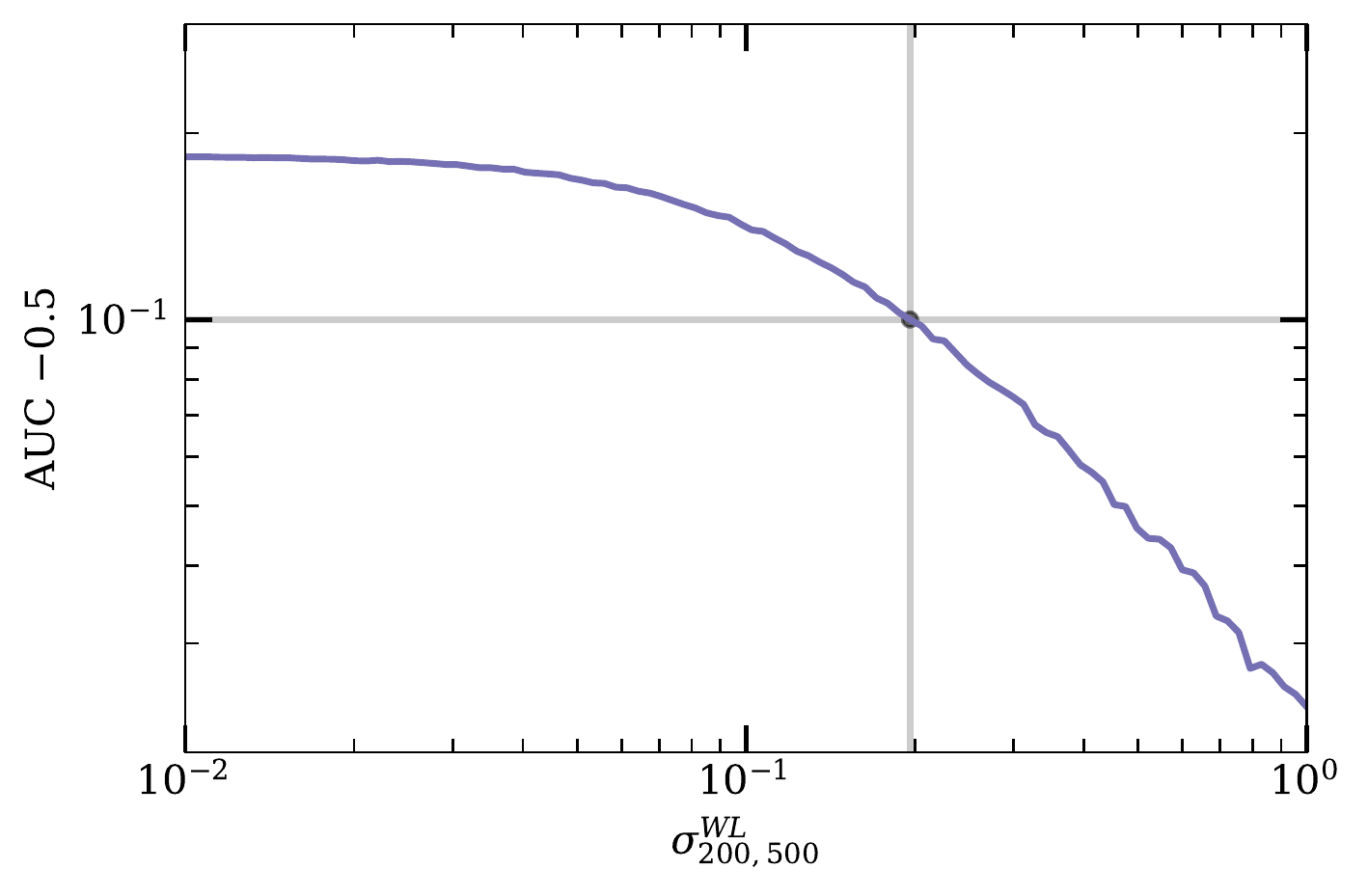}
    \caption{Area Under the ROC Curve (AUC) as function of the scatter on the measured sparsity for WL mass estimates. A random classifier has an AUC$=0.5$. The vertical and horizontal lines denote AUC = 0.6 and the corresponding scatter $\sigma^{\rm WL}_{200,500}=0.2$, denoting the point, $\sigma^\text{WL}_{200,500} > 0.2$, beyond which the detector can be considered ineffective at detecting recent mergers.}
    \label{fig:AUC-scatter}
\end{figure}

\begin{table}
    \centering
    \caption{Sparsity bias and scatter obtained from the weak lensing mass bias estimates by \citet{Becker2011}.}
    \begin{tabular}{cccc}
         \hline
          & $n_\text{gal}$ & $b^\text{WL}_\text{200,500}$ & $\sigma^\text{WL}_\text{200,500}$\\
         \hline
          & $10$ & $0.04\pm0.02$ & $ 0.51\pm0.03 $\\
         $z=0.25$ & $20$ & $ 0.01\pm0.01 $ & $ 0.40\pm0.02 $\\
          & $40$ & $ 0.03\pm0.01 $ & $ 0.35\pm0.02 $\\
          & & &\\
          & $10$ & $0.07\pm0.07$ & $ 0.76\pm0.03 $\\
         $z=0.5$ & $20$ & $ 0.02\pm0.02 $ & $ 0.58\pm0.04 $\\
          & $40$ & $ 0.03\pm0.01 $ & $ 0.49\pm0.03 $\\
          \hline
    \end{tabular}
    
    \label{tab:WL_bias}
\end{table}
We may notice that although the deterministic sparsity bias is smaller than that on individual mass estimates the scatter can be large. In order to evaluate the impact of such biasses on the identification of merging clusters using sparsity estimates we use the values of the bias parameters quoted in Tab.~\ref{tab:WL_bias} to generate a population of biased sparsities using Eq.~(\ref{spars_wl_bias}) with the constraint that $s_{200,500}^\text{WL} > 1$ for our validation sample at $z=0.25$. We then performed the frequentist test for a single sparsity measurements (the Bayesian estimator has a detection power similar to that of the frequentist one.) and evaluated the Area Under the ROC curve (AUC) as function of the scatter $\sigma^{\rm WL}_{200,500}$ to quantify the efficiency of the estimator at detecting recent major merger events. This is shown in Fig.~\ref{fig:AUC-scatter}. Notice that a classifier should have values of AUC$>0.5$ \citep{Fawcett2006}. Hence, we can see that the scatter can greatly reduce the detection power of the sparsity estimator and render the method ineffective at detecting recent mergers for $\sigma^{\rm WL}_{200,500}>0.2$. In contrast, the estimator is valuable classifier for smaller values of the scatter.

\subsubsection{Hydrostatic Mass Bias}
Measurements of galaxy cluster masses from X-ray observations rely on the hypothesis that the intra-cluster gas is in hydrostatic equilibrium. Deviations from this condition can induce a radially dependent bias on the cluster masses \citep[see e.g.][]{2016ApJ...827..112B,Eckert2019,Ettori2022},  thus affecting the estimation of the cluster's sparsity. The hydrostatic mass bias has been studied in \citet{2016ApJ...827..112B}, who have realised cosmological zoom N-body/hydro simulations of 29 clusters to evaluate the bias of masses at overdensities $\Delta=200, 500$ and $2500$ (in units of the critical density) for Cool Core (CC) and No Cool Core (NCC) clusters, as defined with respect to the entropy in the core of their sample, as well as for Regular and Disturbed clusters defined by the offset of the centre of mass and the fraction of substructures.

\begin{table}
    \centering
    \caption{Sparsity bias from the hydrostatic mass bias estimated of \citet{2016ApJ...827..112B} for different categories of simulated clusters.}
    \begin{tabular}{lccc}
        \hline
          & $b_{200,500}^\text{HE}$ & $b_{500,2500}^\text{HE}$ & $b_{200,2500}^\text{HE}$ \\
        \hline
        All & $0.003\pm0.032$ & $-0.037\pm0.025$ & $-0.033\pm0.034$ \\
        CC & $-0.009\pm0.031$ & $-0.151\pm0.038$ & $-0.162\pm0.041$ \\
        NCC & $0.019\pm0.046$ & $0.005\pm0.027$ & $0.023\pm0.041$ \\
        Regular & $0.032\pm0.089$ & $0.025\pm0.037$ & $0.057\pm0.082$ \\
        Disturbed & $-0.017\pm0.077$ & $-0.080\pm0.086$ & $-0.098\pm0.052$\\
        \hline
    \end{tabular}
    \label{tab:hydro_biasses}
\end{table}

Following the evaluation presented in \citet{Corasaniti2018}, we use the hydrostatic mass bias estimates given in Tab.~1 of \citet{2016ApJ...827..112B} to estimate the bias on cluster sparsities, these are quoted in Tab.~\ref{tab:hydro_biasses}. Overall, we can see that the hydrostatic mass bias does not significantly affect the estimated sparsity, with a bias of the order of few percent and in most cases compatible with a vanishing bias with only a few exceptions. This is consistent with the results of the recent analysis based on observed X-ray clusters presented in \citet{Ettori2022}, which yield sparsity biasses at the percent level and consistent with having no bias at all. However, we have seen in the case of the WL mass bias that even though the effect on the measured sparsity remains small, the scatter around the true sparsity can severely affect the efficiency of the detector at identifying recent mergers. Unfortunately, the limited sample from \citet{2016ApJ...827..112B} does not allow to compute the hydrostatic mass bias scatter of the sparsity. If the latter behaves in the same manner as in the WL case, then we can expect the estimator to respond to the increasing scatter as in Fig.~\ref{fig:AUC-scatter}. Consequently, as long as the scatter remains small, $\sigma^{\rm HE}_{\Delta_1,\Delta_2} < 0.1$, then the efficiency of the estimator will remain unaffected.

\subsubsection{Concentration Mass Bias}

We have seen in Section~\ref{sparsprof} that sparsities deduced from the concentration parameter of a NFW profile fitted to the halo density profile are biased compared to those measured using N-body masses. In particular, as seen in Fig.~\ref{fig:relative_spars_conc}, concentration deduced sparsities tend to underestimate their N-body counterparts. Hence, they are more likely to be associated with relaxed clusters than systems in a perturbed state characterised by higher values. A notable exception is the case of haloes undergoing recent mergers which are associated to lower concentration values, or equivalently higher sparsity, even though the N-body estimated sparsity is low. This effect is most likely due to poor fit agreement \citep{Balmes2014}, and systematically increases the population of perturbed haloes above the detection threshold. The concurrences of these two effects leads to an apparent increase in detection power for the 1S estimators when using NFW-concentration estimated masses, as can be seen for the solid lines in Fig.~\ref{fig:validation_roc_curves}.

In contrast when looking at the 3S case in Fig.~\ref{fig:validation_roc_curves}, there is a clear decrease in the detection power for the concentration based sparsity estimates. This is due to the differences in the pulse patterns deduced from concentration compared to the direct measurement of the sparsity, which results in a shape of the pulse at inner radii that is significantly different from that obtained using the N-body masses. Similarly to the 1S estimator, the sparsities measured using the NFW concentration are on average shifted towards smaller values. As such, the effect of using concentration based estimates results in an overestimation of the likelihood that a halo has not undergone a recent merger.

Keeping the above discussions in mind we now present example applications to two well studied galaxy clusters.

\subsection{Abell 383}
Abell 383 is a cluster at $z=0.187$ that has been observed in X-ray \citep{2004A&A...425..367B,2006ApJ...640..691V} and optical bands \citep{2002PASJ...54..833M,2012ApJS..199...25P} with numerous studies devoted to measurements of the cluster mass from gravitational lensing analyses \citep[e.g.][]{2016MNRAS.461.3794O,2016ApJ...821..116U,2019MNRAS.488.1704K}. The cluster appears to be a relaxed system  with HE masses $M_{500\text{c}}=(3.10\pm 0.32)\cdot 10^{14}\,\text{M}_{\odot}$ and $M_{2500\text{c}}=(1.68\pm 0.15)\cdot 10^{14}\,\text{M}_{\odot}$ from Chandra X-ray observations \citep{2006ApJ...640..691V}, corresponding to the halo sparsity $s_{500,2500}=1.84\pm 0.25$ that is close to the median of the halo sparsity distribution. We compute the merger test statistics of Abell 383 using the lensing masses estimates from the latest version of the Literature catalogues of Lensing Clusters \citep[LC$^2$][]{2015MNRAS.450.3665S}. In particular, we use the mass estimates obtained from the analysis of the latest profile data of \citep{2019MNRAS.488.1704K}: $M_{2500\text{c}}=(2.221\pm 0.439)\cdot 10^{14}\,\text{M}_{\odot}$, $M_{500\text{c}}=(5.82\pm 1.15)\cdot 10^{14}\,\text{M}_{\odot}$ and $M_{200\text{c}}=(8.55\pm 1.7)\cdot 10^{14}\,\text{M}_{\odot}$. These give the following set of sparsity values: $s_{200,500}=1.47\pm 0.41$, $s_{200,2500}=3.85\pm 1.08$ and $s_{500,2500}=2.62\pm 0.73$.  We obtain a p-value ${\rm p}=0.21$ and Bayes Factor $B_\text{f}=0.84$, incorporating errors on the measurement of $s_{200,500}$ yields a higher p-value, ${\rm p}=0.40$, which can be interpreted as an effective sparsity of $s^\text{eff}_{200,500} = 1.40$. These results disfavour the hypothesis that the cluster has gone through a major merger in its recent history.

\subsection{Abell 2345}
\begin{figure}
    \centering
    \includegraphics[width = 0.9\linewidth]{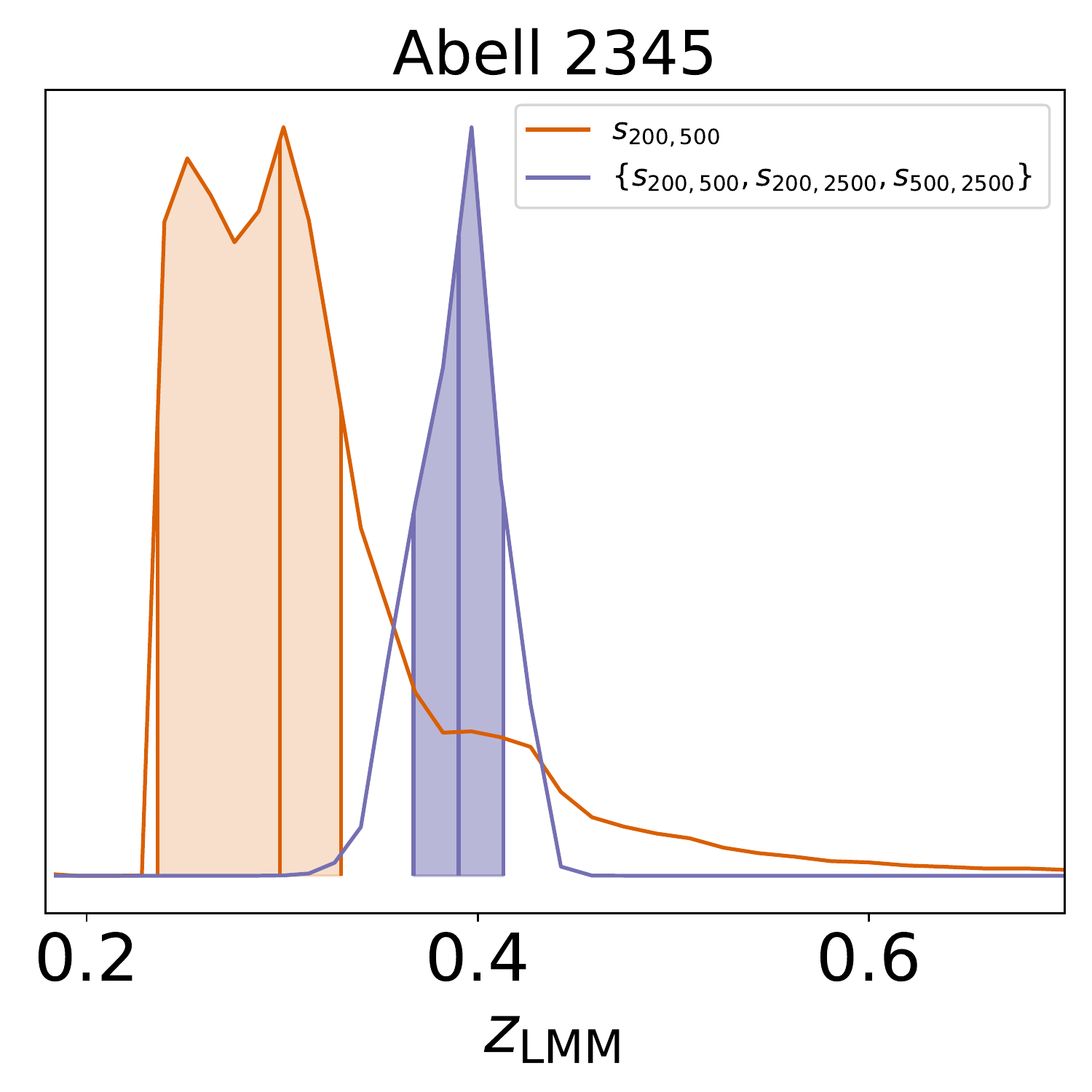}
    \caption{Posterior distributions Abell 2345 obtained using three sparsity measurements from the lensing cluster masses in the LC$^2$ catalogue \citep{2015MNRAS.450.3665S} using the shear data from \citep{2010PASJ...62..811O}. The vertical lines indicates the median value of $z_{\rm LMM}$, while the shaded are corresponds to the $68\%$ credible region around the median.}
    \label{fig:post_A2345}
\end{figure}

Abell 2345 is a cluster at $z=0.179$ that has been identified as a perturbed system by a variety of studies that have investigated the distribution of the galaxy members in optical bands \citep{2002ApJS..139..313D,2010A&A...521A..78B} as well as the properties of the gas through radio and X-ray observations \citep[e.g.][]{1999NewA....4..141G,2009A&A...494..429B,2017ApJ...846...51L,2019ApJ...882...69G,2021MNRAS.502.2518S}. The detection of radio relics and the disturbed morphology of the gas emission indicate that the cluster is dynamically disturbed. Furthermore, the analysis by \citet{2010A&A...521A..78B} suggests that the system is composed of three sub-clusters. \citet{2002ApJS..139..313D} have conducted a weak lensing study on a small field of view centred on the main sub-cluster and found that the density distribution is roughly peaked on the bright central galaxy. This is also confirmed by the study of \citet{2004ApJ...613...95C}, however the analysis by \citet{2010PASJ...62..811O} on a larger field-of-view has indeed shown that Abell 2345 has a complex structure. The shear data have been re-analysed to infer lensing masses that are reported in latest version the LC$^2$-catalogue \citep{2015MNRAS.450.3665S}: $M_{200\text{c}}=(28.44\pm 10.76)\cdot 10^{14}\,\text{M}_{\odot}$, $M_{500\text{c}}=(6.52\pm 2.47)\cdot 10^{14}\,\text{M}_{\odot}$ and $M_{2500\text{c}}=(0.32\pm 0.12)\cdot 10^{14}\,\text{M}_{\odot}$. These mass estimates give the following set of sparsity values: $s_{200,500}= 4.36\pm 2.33$, $s_{200,2500}=87.51\pm 46.83$ and $s_{500,2500}=20.06\pm 10.74$. Using only the $s_{200,500}$ estimate result in a very small p-value, ${\rm p}=4.6\cdot 10^{-5}$. Incorporating errors on the measurement of $s_{200,500}$ yields a higher p-value, ${\rm p}=7.5\cdot10^{-4}$, which can be interpreted as an effective sparsity of $s^\text{eff}_{200,500} = 2.76$, significantly lower than the measured value, however both strongly favour the signature of a major merger event, that is confirmed by the combined analysis of the three sparsity measurements for which we find a divergent Bayes factor. In Fig.~\ref{fig:post_A2345} we plot the marginal posterior for the single sparsity $s_{200,500}$ (orange solid line) and for the ensemble of sparsity estimates (purple solid line). In the former case with obtain a median redshift $z_{\rm LMM}=0.30^{+0.03}_{-0.06}$, while in the latter case we find $z_\text{LMM} = 0.39\pm 0.02$, which suggests that a major merger event occurred $t_\text{LMM} = 2.1\pm 0.2$ Gyr ago. One should however note that in light of the discussions presented above, this result could be associated to a more recent merger event which, as can be seen in Fig.~\ref{fig:test_metrics}, are artificially disfavoured by our method.

\section{Conclusions}\label{conclu}
In this work we have investigated the properties of the mass profile of massive dark matter haloes hosting galaxy clusters. We have focused on haloes undergoing major merger events with the intent of finding observational proxies of the halo mass distribution that can provide hints of recent mergers in galaxy clusters. To this purpose we have performed a thorough analysis of N-body halo catalogues from the MultiDark-Planck2 simulation. 

We have shown that halo sparsity provides a good proxy of the halo mass profile, especially in the case of merging haloes whose density profile significantly deviates from the NFW formula. We have found that major mergers leave a characteristic universal imprint on the evolution of the halo sparsity. This manifests as a rapid pulse response to the major merger event with a shape that is independent of the time at which the major merger occurs. The onset of the merger systematically increases the value of the sparsity, suggesting that mass in the inner part of the halo is displaced relative to the mass in the external region. Following the pulse the value of the sparsity, a quiescent evolution of the halo mass distribution is recovered within only $\sim 2$ dynamical times, which is consistent with the findings of the concentration analysis by \citet{Wang2020}.

The universal imprint of major mergers on the evolution of halo sparsity implies the universality of the distribution of halo sparsities of merging and quiescent haloes respectively. That is to say that at any given redshift it is possible to distinctly characterise the distribution of merging and quiescent haloes. This is because the distribution of sparsity values of haloes that have undergone their last  major merger within $|T|\lesssim 2$ dynamical times differs from that of quiescent haloes that had their last major merger at earlier epochs, $|T|\gtrsim 2$. The former constitutes a sub-sample of the whole halo population that largely contributes to the scatter of the halo sparsity distribution with their large sparsity values. 

The characterisation of these distributions enable us to devise statistical tests to evaluate whether a cluster at a given redshift and with given sparsity estimates has gone through a major merger in its recent history and eventually at which epoch. To this purpose we have developed different metrics based on a standard binary frequentist test, Bayes Factors and Support Vector Machines. We have shown that having access to cluster mass estimates at three different overdensities, allowing to obtain three sparsity estimates, provides more robust conclusions. In the light of these results we have developed a numerical code that can be used to investigate the presence of major mergers in observed clusters. As an example case, we have considered Abell 2345 a known perturbed clusters as well as Abell 383 a known quiescent cluster. 

In the future we plan to expand this work in several new directions. On the one hand, it will be interesting to assess the impact of baryons on halo sparsity estimates especially for merging haloes. This should be possible through the analysis of N-body/hydro simulations of clusters. On the other hand, it may be also useful to investigate whether the universality of the imprint of major mergers on the evolution of halo sparsity depends with the underlying cosmological model. The analysis of N-body halo catalogues from simulations of non-standard cosmological scenarios such as the RayGalGroupSims suite \citep{Corasaniti2018,2021arXiv211108745R}, may allow us to address this point. 

It is important to stress that the study presented here focuses on the statistical relation between halo sparsity and the epoch of last major merger defined as the time when the parent halo merges with a smaller mass halo that has at least one third of its mass. This is different from the collision time, or the central passage time of two massive haloes, which occur on a much shorter time scale. Hence, the methodology presented here cannot be applied to Bullet-like clusters that have just gone through a collision, since the distribution of the collisionless dark matter component in the colliding clusters has not been disrupted and their merger has yet to be achieved. Overall, our results opens the way to timing major merger in perturbed galaxy clusters through measurements of dark matter halo sparsity.

\section*{Acknowledgements}
We are grateful to Stefano Ettori, Mauro Sereno and the anonymous referee for carefully reading the manuscript and their valuable comments. 

The CosmoSim database used in this paper is a service by the Leibniz-Institute for Astrophysics Potsdam (AIP).
The MultiDark database was developed in cooperation with the Spanish MultiDark Consolider Project CSD2009-00064. 
The authors gratefully acknowledge the Gauss Centre for Supercomputing e.V. (www.gauss-centre.eu) and the Partnership for Advanced Supercomputing in Europe (PRACE, www.prace-ri.eu) for funding the MultiDark simulation project by providing computing time on the GCS Supercomputer SuperMUC at Leibniz Supercomputing Centre (LRZ, www.lrz.de).

We thank Instituto de Astrofisica de Andalucia (IAA-CSIC), Centro de Supercomputacion de Galicia (CESGA) and the Spanish academic and research network (RedIRIS) in Spain for hosting Uchuu DR1 in the Skies \& Universes site for cosmological simulations. The Uchuu simulations were carried out on Aterui II supercomputer at Center for Computational Astrophysics, CfCA, of National Astronomical Observatory of Japan, and the K computer at the RIKEN Advanced Institute for Computational Science. The Uchuu DR1 effort has made use of the skun@IAA\_RedIRIS and skun6@IAA computer facilities managed by the IAA-CSIC in Spain (MICINN EU-Feder grant EQC2018-004366-P).
%%%%%%%%%%%%%%%%%%%%%%%%%%%%%%%%%%%%%%%%%%%%%%%%%%
\section*{Data Availability}

During this work we have used publicly available data from the MDPL2 simulation suite \citep{Klypin2016}, provided by the CosmoSim database \href{https://www.cosmosim.org/}{https://www.cosmosim.org/}, in conjunction with publicly available data from the Uchuu simulation suite \citep{Ishiyama2021}, provided by the Skies and Universes database \href{http://skiesanduniverses.org/}{http://skiesanduniverses.org/}.

The numerical code \textsc{lammas} used for this analysis are available at: \href{https://gitlab.obspm.fr/trichardson/lammas}{https://gitlab.obspm.fr/trichardson/lammas}. The package also contains the detailed fitting parameters of the 1S and 3S likelihood distributions for all Uchuu snapshots up to z = 2.



\bibliographystyle{mnras}
\bibliography{bibliography} % if your bibtex file is called example.bib

%\begin{thebibliography}{99}
%\end{thebibliography}

%%%%%%%%%%%%%%%%%%%%%%%%%%%%%%%%%%%%%%%%%%%%%%%%%%

%%%%%%%%%%%%%%%%% APPENDICES %%%%%%%%%%%%%%%%%%%%%

\appendix

\section{LAMMAS Code}\label{LAMMAS}
We have developed a specialised analysis pipeline \textsc{lammas} that implements the steps detailed throughout this work. This Python library, based on the \textsc{emcee} \citep{Emcee2013} library, is designed to estimate the posterior distributions and test statistics presented in this work to allow one to estimate whether a cluster with given sparsity measurements has gone through a major merger and eventually when such event has occurred.

As an input the user specifies the mass measurements of the galaxy cluster under investigation, $M_{200\text{c}}$ and $M_{500\text{c}}$ which are mandatory, and $M_{2500\text{c}}$ which is optional. In addition, the user is required to input the redshift of the cluster and the assumed cosmology, although for the time being the cosmology must be set to that of the numerical simulation used in the calibration, namely the Planck cosmology of the Uchuu simulations. From this information the code then carries out a MCMC estimation of the posterior distribution using 1 and 3 sparsities if $M_{2500\text{c}}$ has been specified. This is done using a Metropolis Hastings algorithm over a total of $10^5$ steps by default. The likelihood functions used in these estimations are those of Eq.~(\ref{eq:gen_beta_prime}) and Eq.~({\ref{eq:like3D}}) respectively for the 1 and 3 sparsity estimates. The redshift dependence of the parameters of these distributions is taken into account by interpolating tabulated measurements calibrated using simulation data.

Priors are taken into account during the posterior estimation. By default the program uses a uniform prior in $a_\text{LMM}$ as used throughout this work, but the user can instead choose to use uniform priors in $z_\text{LMM}$ or $t_\text{LMM}$, all being distinct from one another. At the time of writing more complex prior selection has not been implemented.

The resulting output takes the form of a python dictionary containing posterior distributions, median likelihoods and corresponding credible intervals but also the $p-$value corresponding to the measurement of $s_{200,500}$ and Bayes factors corresponding to the 1 and 3 sparsity estimates.

%%%%%%%%%%%%%%%%%%%%%%%%%%%%%%%%%%%%%%%%%%%%%%%%%%

% Don't change these lines
\bsp	% typesetting comment
\label{lastpage}
\end{document}